\documentclass[10pt,a4paper]{article}
\usepackage{amsmath,amssymb,graphicx,subfigure,geometry,indentfirst,textgreek,natbib,aas_macros,color}
\DeclareMathOperator{\erfc}{erfc}
\begin{document}
\title{A Three-component Model for Cosmic-ray Spectrum and Dipole Anisotropy}
\author{Yiran Zhang\footnote{Key Laboratory of Planetary Sciences, Purple Mountain Observatory, Chinese Academy of Sciences, Nanjing 210023, China (zhangyr@pmo.ac.cn)}, Siming Liu\footnote{School of Physical Science and Technology, Southwest Jiaotong University, Chengdu 610031, China (liusm@swjtu.edu.cn)}, and Houdun Zeng\footnote{Key Laboratory of Dark Matter and Space Astronomy, Purple Mountain Observatory, Chinese Academy of Sciences, Nanjing 210023, China (zhd@pmo.ac.cn)}\ \footnote{Key Laboratory of Astroparticle Physics of Yunnan Province, Kunming, 650091, China}}
\maketitle
\begin{abstract}
Using a three-component, multi-scale diffusion model, we show that the cosmic-ray (CR) proton and helium spectra and the dipole anisotropy can be explained with reasonable parameters. The model includes a nearby source associated with the supernova remnant (SNR) that gave rise to the Geminga pulsar, a source at the Galactic center, and a component associated with the Galactic disk. The CR flux below TeV is dominated by the disk component. The center source with a continuous injection of CRs starting about 18 Myr ago is needed to explain the anisotropy above 100 TeV. With the assumption of universal CR spectra injected by all SNRs, the nearby source can produce a TeV spectral bump observed at Earth via slow diffusion across the interstellar magnetic field, which needs to have an angle $ \theta \approx 5^{\circ} $ between the field line and the line of sight toward the source, and have weak magnetic turbulence with the Alfv\'{e}n Mach number $ M_{\text{A}}\approx 0.1 $. Considering the modulation of the Galactic-scale anisotropy by this magnetic field, in a quasi-local approach the field may be directed at a right ascension about $ -90^{\circ} $ and a declination about $ -7.4^{\circ} $ in the equatorial coordinate system.

\end{abstract}
\section{Introduction}
Recent precise measurements of cosmic rays (CRs), e.g., CREAM-III \citep{2017ApJ...839....5Y}, NUCLEON \citep{2018JETPL.108....5A} and DAMPE \citep{2019SciA....5.3793A,2021PhRvL.126t1102A}, show strong evidence for a bump-like shape of the nucleus energy spectrum in the log-log plot in the energy range of TeV--PeV. This feature carries valuable information about acceleration and propagation characteristics of Galactic CRs, is the key to further investigation of the CR origin problem.

In the typical Galactic propagation scenario, CRs accelerated by a source, which is probably the diffusive shock wave of a supernova remnant (SNR), have a power-law energy spectrum with a high-energy cutoff corresponding to the maximum acceleration ability. It is believed that the Galactic CRs are injected by a large-scale population of sources with similar acceleration spectral indices, and their propagation can reach the steady state of an injection-escape balance. The diffusive escape leads to softening of the injection spectrum by an additional power index reflecting the energy dependence of the diffusion coefficient.

It has been shown that the observed CR spectral hardening at hundreds of GeV can be ascribed to changes of the diffusion property \citep{2012ApJ...752L..13T,2012PhRvL.109f1101B,2017PhRvD..95b3001T,2018ApJ...869..176L}, although existing measurements, e.g., AMS-02 \citep{2016PhRvL.117w1102A} and CREAM \citep{2008APh....30..133A}, seem still insufficient to strictly constrain the energy dependence of the flux ratio of secondary-to-primary nuclei. On the other hand, the spectral hardening may also be reproduced with some detailed modeling of the source distribution \citep{2011ApJ...729L..13O,2011PhRvD..84d3002Y,2015ApJ...815L...1T} and the particle acceleration process \citep{2013ApJ...763...47P,2014A&A...567A..33T}. To some extent, most of these models may further be adjusted to fit the TeV spectral bump. It is remarkable that the time-dependent particle acceleration by a shock with turbulent diffusion can naturally include a spectral softening gradually toward high energies, which can be used to fit the slightly convex spectrum \citep{2017ApJ...844L...3Z}.

However, within a global diffusion model (i.e., assuming a (quasi-)continuous distribution of Galactic CR sources), even the observed amplitude of the CR large-scale anisotropy may be explained, the change of the dipole anisotropy direction (phase) with the CR energy can not be reproduced \citep{2017PrPNP..94..184A}. As reported by observations via extended air showers \citep{2017ApJ...836..153A,2018ApJ...861...93B} and atmospheric muons \citep{2016ApJ...826..220A}, this change is roughly from the Galactic anti-center to center direction as the energy increases from tens to hundreds of TeV, with a closely correlated dip in the log-log plot of the dipole amplitude versus the energy. The coincident energy range of the spectral bump and the anisotropy phase flip gives strong hints for the existence of a (several) nearby source(s) with prominent contribution to the CR intensity observed at Earth \citep{2017PhRvD..96b3006L,2019JCAP...12..007Q,2020arXiv200715321F}.

Perhaps because each of the ground-based observatories can not operate a full-sky scan, they usually do not strictly constrain the declination (Dec), but only report a right-ascension (RA) projection of the dipole anisotropy. This results in a large uncertainty for determining the Galactic longitude of the nearby source \citep{2016PhRvL.117o1103A}. Nevertheless, looking at the reported celestial anisotropy map, it should be plausible that the direction of the TeV CR intensity excess roughly coincides with that of the Geminga pulsar \citep{2008A&A...485..527S}, whose $ \left( \text{RA},\text{Dec} \right) \approx \left( 98.48,17.77 \right) ^{\circ} $.

This pulsar may currently be located at a distance of $ 250^{+120}_{-62} $ pc to Earth. With the ``usual'' order of magnitude $ 10^{29}\text{ cm}^2\text{s}^{-1} $ of the TeV diffusion coefficient in the interstellar medium (ISM), the CRs can sufficiently diffuse from the Geminga, as well as from its possible birth place, to the Earth in a characteristic pulsar age of 340 kyr. In such a diffusion scenario, if a Geminga-like nearby source is responsible for the spectral bump, it should be a unique source with a harder injection spectrum and a lower highest energy compared with Galactic background sources \citep{2019FrPhy..1524601Y}. On the other hand, if the nearby source is similar to the background, our Earth should stand at the diffusion distance of TeV CRs from the source for producing a low-energy cutoff corresponding to the bump, which means that these CRs need to diffuse much slower than those from the background \citep{2017Sci...358..911A}.

A natural connection between the small- and large-scale diffusion property may be based on the fact that CRs are kinematically trapped along the magnetic field, i.e., observers can see highly anisotropic diffusion on scales smaller than the radius of curvature of the field lines, but see isotropic diffusion averaged over large scales due to wandering of the field. It has been shown that the TeV $ \gamma $-ray halo can be explained with the anisotropic diffusion in the vicinity of the pulsar wind nebula \citep[PWN;][]{2019PhRvL.123v1103L}. In this paper, we shall assume that there is a regular magnetic field in the Geminga-to-Earth ISM. The cross-field slow diffusion results in a TeV low-energy cutoff of the CR spectral component from the Geminga, which is partially responsible for the bump in the total spectrum. To simplify the large-scale diffusion, in view of distribution characteristics of Galactic CR sources, we shall also introduce spectral components from the Galactic center and disk. The center component also contributes to the bump, and is the reason for the anisotropy phase flip. The disk component dominates the sub-TeV intensity. We shall also consider the modulation of the anisotropy by the magnetic field with the projection approach \citep{2016PhRvL.117o1103A}.
\section{Multi-scale Diffusion}
As mentioned previously, we try to model the CR diffusion property as simple as possible. Put the diffusion tensor
\begin{align}
\kappa _{ij}=\kappa _{\bot}\left( \delta _{ij}-\frac{B_iB_j}{B^2} \right) +\kappa _{\parallel}\frac{B_iB_j}{B^2}+\kappa _{\text{T}}\varepsilon _{ijk}\frac{B^k}{B},\label{kappa}
\end{align}
where $ \delta _{ij} $ is the Kronecker delta, $ \varepsilon _{ijk} $ is the Levi-Civita symbol, $ \boldsymbol{B} $ is the magnetic field, $ \kappa _{\bot} $ and $ \kappa _{\parallel} $ are the perpendicular and parallel diffusion coefficient from the pitch-angle scattering process, satisfying $ \kappa _{\bot}=M_{\text{A}}^{4}\kappa _{\parallel} $ and $ \kappa _{\parallel}\propto vR^{\varDelta} $, where $ v $ and $ R $ are the speed and rigidity of a charged particle, respectively. The Alfv\'{e}n Mach number $ M_{\text{A}}<1 $ corresponds to the slow diffusion $ \kappa _{\bot}\ll \kappa _{\parallel} $ across the field \citep{2019PhRvL.123v1103L}. For sub-TeV CRs, the boron-to-carbon (B/C) flux ratio measured by the AMS-02 experiment suggests $ \varDelta =1/3 $, which is consistent with the Kolmogorov picture of turbulence \citep{2016PhRvL.117w1102A}. The antisymmetric part of $ \kappa _{ij} $ is introduced due to the density-gradient transverse drift \citep{1975Ap&SS..32...77F}, which, however, leads at most to a convection term instead of a typical diffusion term (i.e., the divergence of the diffusion flux) in the transport equation.

Generally, the diffusion equation can be written as $ \partial f/\partial t=\partial \left( \kappa _{ij}\partial f/\partial x_j \right) /\partial x_i $, where $ f $ is the distribution function, $ x_i $ is the $ i $th component of the position vector $ \boldsymbol{r} $. If $ \kappa _{ij} $ is spatially homogeneous, in a small-scale system with a uniform magnetic field, one has the anisotropic diffusion equation
\begin{align}
\frac{\partial f}{\partial t}=\frac{\kappa _{\bot}}{\rho}\frac{\partial}{\partial \rho}\left( \rho \frac{\partial f}{\partial \rho} \right) +\kappa _{\parallel}\frac{\partial ^2f}{\partial \zeta ^2},\label{aniso-diff}
\end{align}
where $ \rho $ and $ \zeta $ are spatial coordinates perpendicular and parallel to the magnetic field, respectively. On a sufficiently large scale, the magnetic field may be randomized in its direction, but still be a guide field due to small gyro-radii of particles, which ensures the pitch-angle scattering to be the fundamental process of the diffusion. The statistical average of Eq.~(\ref{kappa}) over the randomized field direction leads to $ \left< \kappa _{ij} \right> =D\delta _{ij} $ with $ D\approx \kappa _{\parallel}/3 $. Considering that any spatial point of the body under consideration is such a ``large-scale'' system, we may take $ \left< f \right> \sim f $, and the isotropic diffusion equation
\begin{align}
\frac{\partial f}{\partial t}=D\nabla ^2f.\label{iso-diff}
\end{align}
Note that, to some extent, the term ``large spatial scale'' is equivalent to ``long evolving time'' according to the ergodic hypothesis.

An impulse point source is described with an additional injection term $ N\delta \left( t\right) \delta ^3\left( \boldsymbol{r} \right) $ in the right-hand side of the diffusion equation, where $ N $ is the (spectrum of) total particle number injected, $ r $ is the distance to the source, $ \delta $ represents the Dirac delta function. For the small-scale system Eq.~(\ref{aniso-diff}) with uniform diffusion, such a source yields the anisotropic three-dimensional (3D) Gauss distribution function
\begin{align}
f=\frac{N}{4\pi \kappa _{\bot}t\sqrt{4\pi \kappa _{\parallel}t}}e^{-\frac{\rho ^2}{4\kappa _{\bot}t}-\frac{\zeta ^2}{4\kappa _{\parallel}t}},\label{fn}
\end{align}
and the diffuse dipole anisotropy in the BGK approximation \citep{2017PrPNP..94..184A}
\begin{align}
A_i=\frac{3\kappa _{ij}}{vf}\frac{\partial f}{\partial x_j}=-\frac{3}{2vt}\left( x_i+\varepsilon _{ijk}\rho ^j\frac{B^k}{B}\frac{\kappa _{\text{T}}}{\kappa _{\bot}} \right) .\label{Ai}
\end{align}
A complete representation of $ \kappa _{\text{T}} $ seems to be complicated \citep{2017ApJ...841..107E}. Moreover, a large value of $ \kappa _{\text{T}} $ (with the typical high-energy limit $ vR/\left( 3B\right) $) may result in a fast increase of the above anisotropy with the particle energy, which is inconsistent with the observation. We thus adopt $ \rho \left| \kappa _{\text{T}}\right| \ll r\kappa _{\bot} $ to drop the $ \kappa _{\text{T}} $ term for simplicity. Then, the diffuse anisotropy always points toward the source and is independent of the uniform magnetic field. Assuming such a regular field in the nearby ISM, the above results can be applied to the CR distribution component produced by the nearby source, i.e., an SNR with its CR acceleration terminated a long time ago.

On the other hand, CRs from Galactic background sources should propagate over a much larger scale than nearby sources, i.e., they propagate with Eq.~(\ref{iso-diff}). These background sources are mainly distributed along the Galactic disk, with an increasing source density toward the Galactic center \citep{2019JCAP...12..007Q}, implying that there are two main components from the background. For the sake of simplicity, we characterize one of them as arising from sources homogeneously distributed along an infinite disk with a surface density $ \mu $ and within a thickness $ 2h $. Imposing a particle escape boundary at a vertical distance $ \left| z \right| =L $ to the Galactic plane $ z=0 $, the steady state $ \partial f/\partial t=0 $ exists under the injection-escape balance, in which each source on average has a particle injection rate $ Q $. Integration of the steady-state diffusion equation gives a quadratic and linear function for $ \left| z \right| <h $ and $ h<\left| z \right| <L $, respectively. The four constants of integration can be determined with conditions of the escape boundary $ f|_{\left| z\right| =L}=0 $, and the smoothness $ f|_{\left| z\right| =h^-}=f|_{\left| z\right| =h^+} $, $ \partial f/\partial z|_{\left| z \right|=h^-}=\partial f/\partial z|_{\left| z \right|=h^+} $, $ \partial f/\partial z|_{z=0}=0 $. For $ h<\left| z \right| <L $, one has $ 2Df=\mu Q\left( L-\left| z \right| \right) $. Since our Earth is currently very close to the Galactic plane, we are more interested in the source region $ \left| z \right| <h $, in which
\begin{align}
f=\frac{L\mu Q}{2D}\left\{ 1-\frac{h}{2L}\left[ 1+\left( \frac{z}{h} \right) ^2 \right] \right\} .\label{fd}
\end{align}
Obviously, on the large scale the average anisotropy is
\begin{align}
\left< \boldsymbol{A}\right> =\frac{3D\boldsymbol{z}}{vfz}\frac{\partial f}{\partial z}=\frac{3D\boldsymbol{z}}{vLh}\left\{ \frac{h}{2L}\left[ \left( \frac{z}{h} \right) ^2+1 \right] -1 \right\} ^{-1}\label{Ad}
\end{align}
with $ \boldsymbol{z} $ the 1D position vector.

We characterize the other background component as arising from a point source located at the Galactic center. Because of the finite escape boundary in the $ z $ direction, the Green function $ G $ can simply be understood with discretization of $ z $-direction eigenvalues, i.e., transferring the Fourier integral of the $ z $-direction Gauss distribution into a Fourier series as follows,
\begin{align}
\frac{1}{\left( 4\pi Dt \right) ^{\frac{3}{2}}}e^{-\frac{r^2}{4Dt}}&=\frac{1}{4\pi Dt}e^{-\frac{\varrho ^2}{4Dt}}\frac{1}{\pi}\int_0^{\infty}{\cos \left( kz \right) e^{-k^2Dt}dk}\notag \\
&\rightarrow \frac{1}{4\pi Dt}e^{-\frac{\varrho ^2}{4Dt}}\frac{1}{L}\sum_{n=0}^{\infty}{\cos \left[ \frac{\pi z}{L}\left( n+\frac{1}{2} \right) \right] e^{-\left[ \frac{\pi}{L}\left( n+\frac{1}{2} \right) \right] ^2Dt}}=G,\label{G}
\end{align}
where $ \varrho $ is the radius in the Galactic plane, and the series can also be expressed in terms of Jacobi theta functions, which are not invoked here due to their scattered definitions. As $ z/L\rightarrow 0 $ and $ Dt/L^2\rightarrow 0 $, the series approaches the integral, leading to the reduction of $ G $ to the typical form of uniform diffusion, which has a spectral low-energy cutoff in accordance with $ r^2\sim 4Dt $. For any $ z<L $ and $ z<\varrho $, the cutoff is evaluated with $ \varrho ^2\sim 4Dt $. For $ Dt/L^2\rightarrow \infty $, only the $ n=0 $ term is important, which gives a high-energy cutoff corresponding to $ L^2\sim \left( \pi /2 \right) ^2Dt $. Therefore, in conclusion, particles with an impulse injection can not effectively propagate out of $ \varrho <L $. On the other hand, in a steadily continuous injection model with the injection term $ Q\delta ^3\left( \boldsymbol{r} \right) $, we obtain
\begin{align}
f&=Q\int_0^t{G\left( t-t' \right) dt'}=\frac{Q}{4\pi DL}\sum_{n=0}^{\infty}{\int_0^{\frac{Dt}{L^2}}{e^{-\left( \frac{\varrho}{2L} \right) ^2\frac{1}{\tau}-\left[ \pi \left( n+\frac{1}{2} \right) \right] ^2\tau}\frac{d\tau}{\tau}}},\label{fc}\\
\left< \boldsymbol{A} \right> &=\frac{3D\boldsymbol{\varrho }}{vf\varrho}\frac{\partial f}{\partial \varrho}=-\frac{3D\boldsymbol{\varrho }}{2vL^2}\frac{\sum_{n=0}^{\infty}{\int_0^{\frac{Dt}{L^2}}{e^{-\left( \frac{\varrho}{2L} \right) ^2\frac{1}{\tau}-\left[ \pi \left( n+\frac{1}{2} \right) \right] ^2\tau}\frac{d\tau}{\tau ^2}}}}{\sum_{n=0}^{\infty}{\int_0^{\frac{Dt}{L^2}}{e^{-\left( \frac{\varrho}{2L} \right) ^2\frac{1}{\tau}-\left[ \pi \left( n+\frac{1}{2} \right) \right] ^2\tau}\frac{d\tau}{\tau}}}},\label{Ac}
\end{align}
where we have taken $ z/L\rightarrow 0 $ for convenience of later use, as it can be shown that here the anisotropy in the $ z $ direction is $ O\left( L/h \right) $ times smaller than that of Eq.~(\ref{Ad}). Similar to the previous analysis, the whole sequence (i.e., starting from $ n=0 $) becomes quasi-continuous as $ Dt/L^2\rightarrow 0 $, for which the series and integral in Eq.~(\ref{fc}) can be expressed with the complementary error function $ \left( L/\varrho \right) \erfc \left( \varrho /\sqrt{4Dt} \right) $, which also contains a low-energy cutoff around $ \varrho ^2\sim 4Dt $. However, the system approaches a steady state as $ Dt/L^2\rightarrow \infty $, in which the integral in Eq.~(\ref{fc}) reduces to the modified Bessel function $ 2K_0\left[ \left( n+1/2 \right) \pi \varrho /L \right] $. In other words, due to the convolution, there is no high-energy cutoff caused by the escape effect of particles in the continuous injection scenario, which will be applied to the center component.

Although at present the large-scale average is the only practical approach to model the observation, it is generally recognized that the anisotropy should be affected by the local magnetic field. However, it is very difficult to reconstruct the ``true'' anisotropy locally observed below the scale on which the statistical average is taken. As implied by Eq.~(\ref{Ai}), the irrotational diffusion flux $ \left( \kappa _{ij}+\kappa _{ji} \right) \left( \partial f/\partial x_j \right) /2 $ is independent of a regular magnetic field. A qualitative description is that the suppression of diffusion results in an increase of the density gradient, which in turn compensates to some extent for the reduction of the diffusion flux. But it seems difficult to determine this compensation in an irregular magnetic field via solving the transport equation because of the lack of detailed information about the Galactic field structure. An in-depth study of the diffuse anisotropy under complicated global diffusion is required, which is beyond the scope of this paper. In the framework of the statistical average, a quasi-local approach may be a simple blend of the large-scale gradient and small-scale diffusion, i.e.,
\begin{align}
A_i=\frac{\kappa _{ij}}{D}\left< A^j \right> \approx 3\frac{B_iB_j}{B^2}\left< A^j \right> \label{Aql}.
\end{align}
In other words, we treat the average gradient as the true gradient contributed to Earth by the large-scale system. We shall also naively try to directly apply the average value $ \left< \boldsymbol{A}\right> $ to fit the observation.
\section{Cosmic-ray Distribution}
If necessary, we shall denote the nearby, Galactic-disk and -center source component by the subscript n, d and c, respectively. For the same type of particles, the total distribution $ f $ observed at the local ISM of the Solar system is given by the scalar superposition, i.e.,
\begin{align}
f=f_{\text{n}}+f_{\text{d}}+f_{\text{c}}.
\end{align}
Note that $ f_{\text{n}} $, $ f_{\text{d}} $, and $ f_{\text{c}} $ are described with Eqs.~(\ref{fn}), (\ref{fd}) and (\ref{fc}), respectively, and will be treated as numbers of particles per unit time, volume and rigidity. We further assume $ N $ and $ Q $ to be power-law rigidity spectra with exponential high-energy cutoff, i.e., to be proportional to $ R^{-\alpha}e^{-R/R_\text{m}} $.

The CR flux $ F $ (per steradian) observed at Earth is modified by the expanding Solar wind. At sufficiently high energies, this modulation may formally be equivalent to a phase-space transformation with the Liouville theorem \citep{1967ApJ...149L.115G,1968ApJ...154.1011G}, i.e.,
\begin{align}
F\left( R \right) =\frac{vR^2}{4\pi R'^2}f\left( R' \right) ,
\end{align}
where $ R' $ and $ R $ refer to rigidities before and after the modulation, respectively. In the force-field approximation, one has $ R'^2=R^2+2R\phi c/v+\phi ^2 $, where $ c $ is the speed of light, $ \phi $ is the modulation potential.

By definition, the total dipole anisotropy $ \boldsymbol{A} $ for a given type of particles should be considered as a weighted average of source components involved, i.e.,
\begin{align}
\boldsymbol{A}=\frac{\boldsymbol{A}_{\text{n}}f_{\text{n}}+\boldsymbol{A}_{\text{d}}f_{\text{d}}+\boldsymbol{A}_{\text{c}}f_{\text{c}}}{f},\label{A}
\end{align}
where $ \boldsymbol{A}_{\text{n}} $ is described with Eq.~(\ref{Ai}) (excluding the $ \kappa_{\text{T}} $ term). As mentioned before, for the magnetic modulation of the background anisotropy, we consider two approaches, one of which, hereafter referred to as the ``naive approach'' for short, simply uses the average value $ \left< \boldsymbol{A}_{\text{d}}\right> $ and $ \left< \boldsymbol{A}_{\text{c}}\right> $ described with Eqs.~(\ref{Ad}) and (\ref{Ac}), respectively. Besides, we also consider the quasi-local approach Eq.~(\ref{Aql}). Note that we can neglect any effect of the force-field Solar modulation on the anisotropy for $ \phi \ll 100\text{ GV} $, because all anisotropy data involved are above 100 GeV. We also ignore any Compton-Getting effect in this analysis.

Obviously, elemental components, in the sense of $ fdR $, should also obey the above law of superposition. For simplicity, we only consider protons (H) and heliums (He) since the energy range of interest is mainly below PeV, which is dominated by these particles. Then, the total dipole anisotropy that needs to be brought into comparison with the observation is $ \left( \boldsymbol{A}_{\text{H}}f_{\text{H}}dR_{\text{H}}+\boldsymbol{A}_{\text{He}}f_{\text{He}}dR_{\text{He}} \right) /\left( f_{\text{H}}dR_{\text{H}}+f_{\text{He}}dR_{\text{He}} \right) $.

Since low-energy CRs may not reach Earth in a finite evolving time, the time-dependent spectrum, i.e., the nearby and center component, should cut off toward low energies. If these cutoffs are responsible for the spectral bump and the anisotropy phase flip in TeV energies, the disk component should be the primary contributor of sub-TeV CRs. Then the large-scale diffusion coefficient $ D $ in our model should be consistent with that in a simple leak-box estimate (see Eq.~(\ref{fd}) for $ h\ll L $). According to the observed sub-TeV B/C flux ratio \citep{2016PhRvL.117w1102A}, we take $ D\left( \text{TV}\right) =10^{29}\text{ cm}^2\text{s}^{-1} $ and $ \varDelta =1/3 $, which correspond to an order of $ 10^{50} $ erg for the total CR energy released by a Galactic SNR-like source in the usual leaky-box approximation. The Fermi-LAT observation of $ \gamma $-ray emissions from interstellar clouds suggests a 50\% decline of CR densities within 2 kpc from the Galactic plane \citep{2015ApJ...807..161T}. Following this result, we should take the escape length $ L=4 $ kpc in our disk model. Note that observational and theoretical studies typically estimate $ L\sim 1 $--10 kpc. Since $ \left| z \right| <h\ll L $, the height of Earth, $ \left| z \right| $, and the half thickness of the source distribution, $ h $, are approximately degenerated into a single parameter $ \left| z/h \right| \lesssim 0.3 $, which has little impact on the spectral fit, but is important to the anisotropy.

The Earth is also located at $ r_\text{c}\approx 8.5 $ kpc from the Galactic center, which is on the order of the large-scale diffusion distance of TeV CRs in a propagation time $ t_\text{c}\sim 20 $ Myr, i.e., $ r_\text{c}^2\sim 4D\left( \text{TV} \right) t_\text{c} $, implying a TeV low-energy cutoff of such a center component. The impulse injection scenario should be excluded for this component due to $ L<\varrho $ (see Eq.~(\ref{G})). Even for $ L\sim 10 $ kpc, CRs diffusing from the impulse center to Earth may still be insufficient because the low- and high-energy cutoff caused by propagation effects are just at comparable TeV energies. We thus consider the continuous injection scenario, which may indicate that the center region is in the midst of an activity that has been ongoing for tens of Myr. In view of $ \left| z \right| \ll L $, Eqs.~(\ref{fc}) and (\ref{Ac}) should be valid approximations with $ \varrho \approx r_\text{c} $.

Since the line-of-sight motion of Geminga is a relatively uncertain factor, we only take $ t_\text{n}=340 $ kyr and $ r_\text{n}=250 $ pc to characterize the location of the nearby source. With such a location, one has $ r_\text{n}^2\ll 4D\left( \text{TV}\right) t_\text{n} $, i.e., the low-energy cutoff in the isotropic diffusion scenario is significantly below the energy range of the bump. In the anisotropic diffusion scenario, a TeV cutoff can be induced by the suppressed diffusion distance perpendicular to the ISM magnetic field even though the parallel diffusion is fast, i.e., $ \rho ^2\sim 4\kappa _{\bot}\left( \text{TV}\right) t_\text{n} $ while $ \zeta ^2\ll 4\kappa _{\parallel}\left( \text{TV}\right) t_\text{n} $.

\begin{figure}[b]
	\centering
	\subfigure[Spectra]{\includegraphics[width=0.496\textwidth]{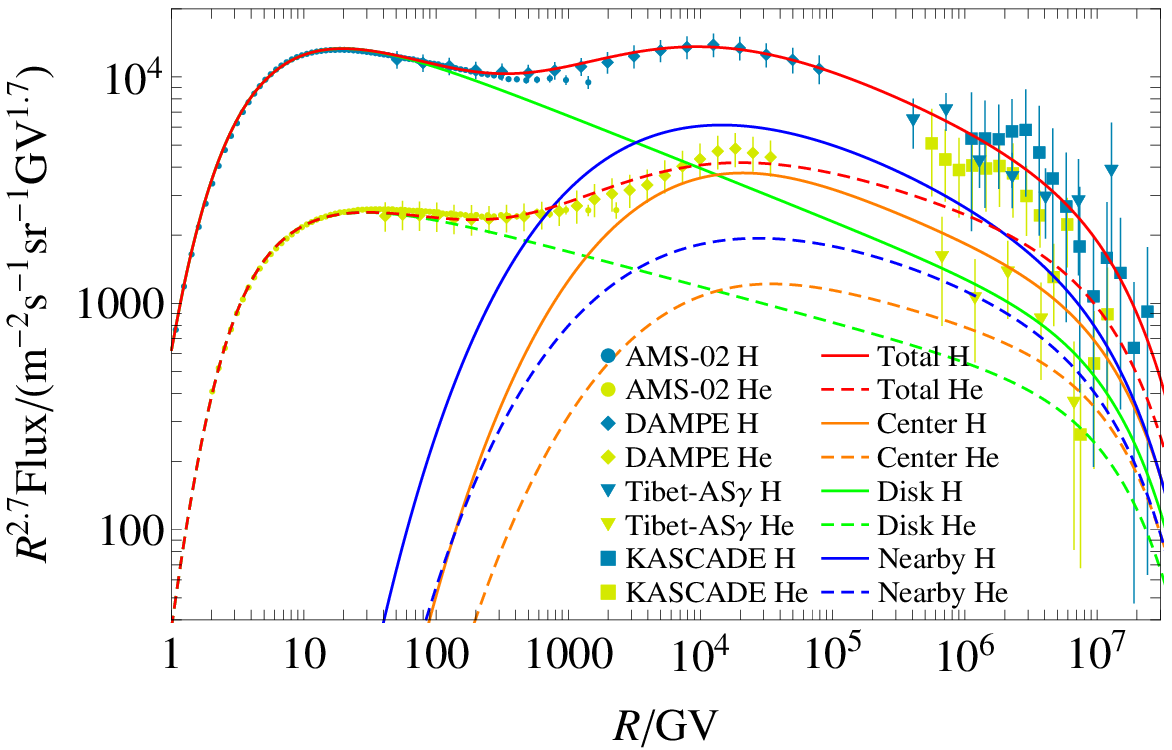}\label{f1a}}
	\subfigure[Dipole Anisotropy Amplitude]{\includegraphics[width=0.496\textwidth]{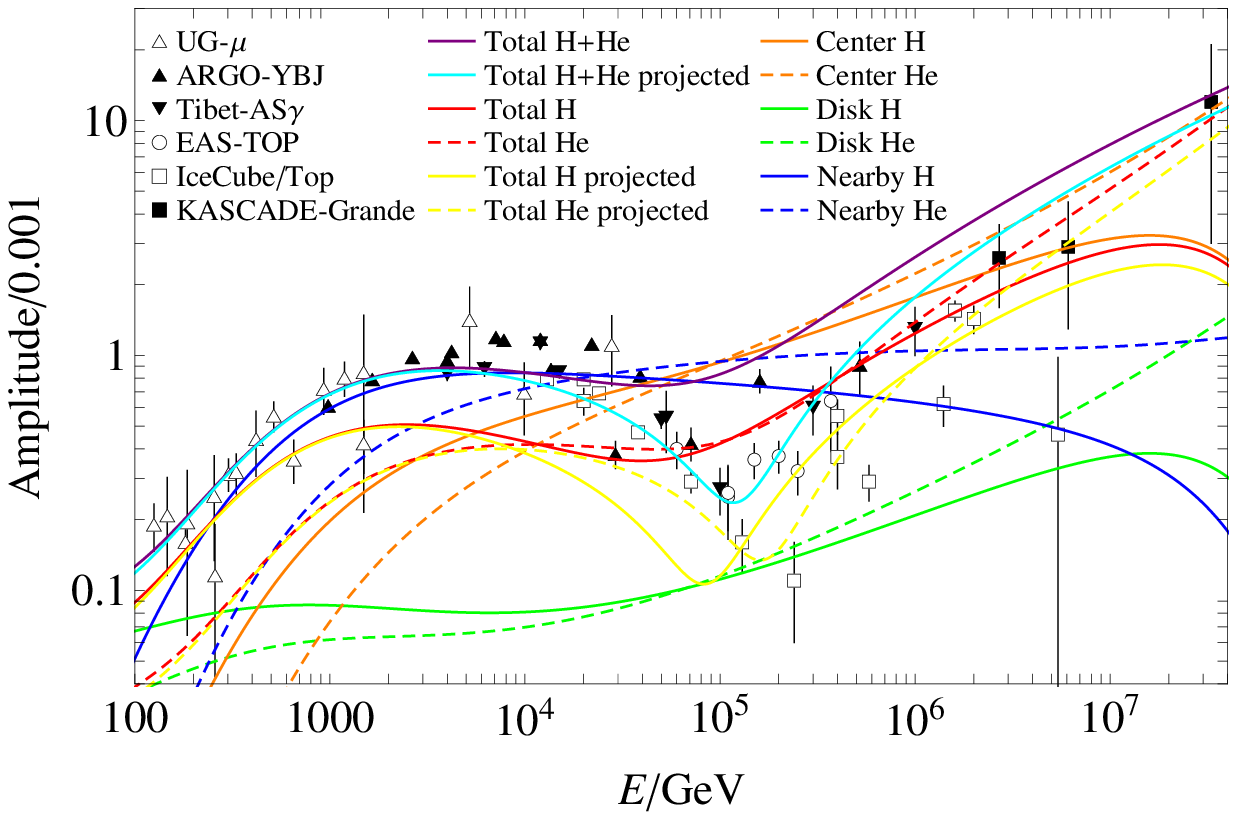}\label{f1b}}
	\subfigure[Dipole Anisotropy Phase]{\includegraphics[width=0.496\textwidth]{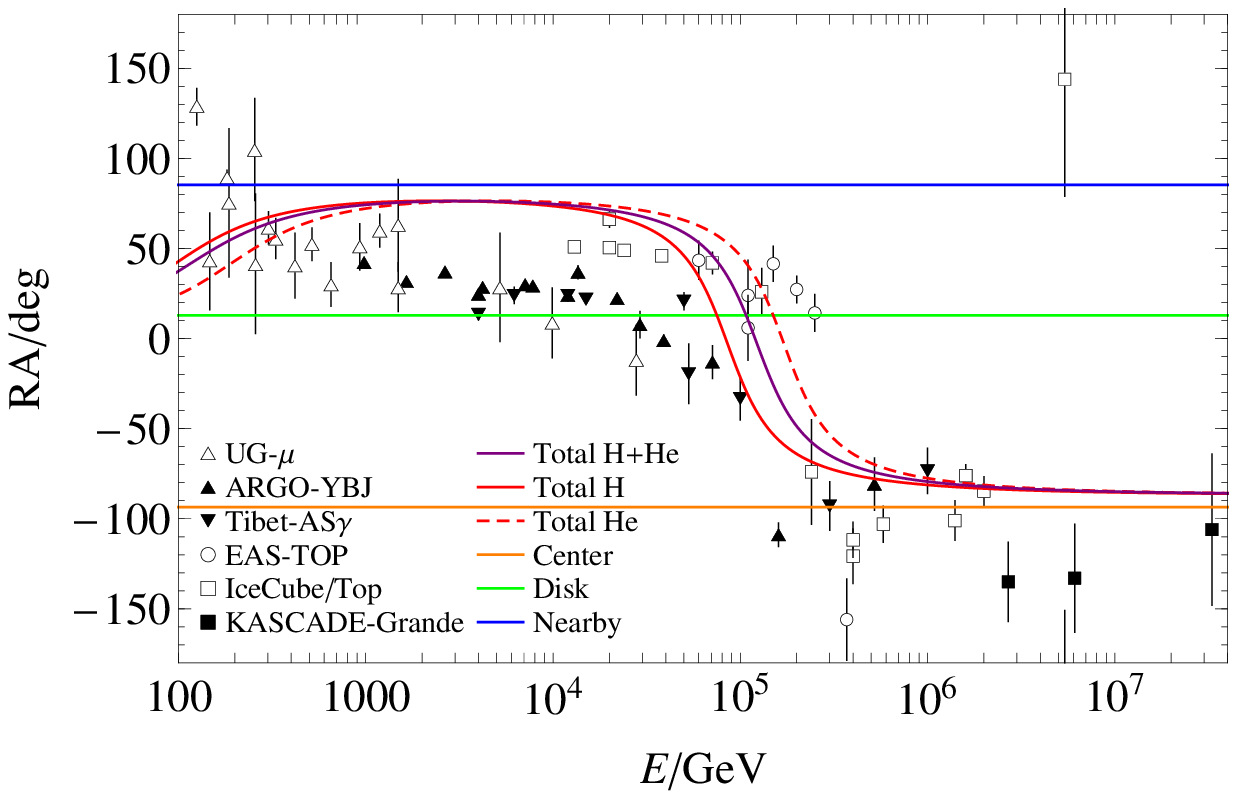}\label{f1c}}
	\subfigure[Sky Map]{\includegraphics[width=0.496\textwidth]{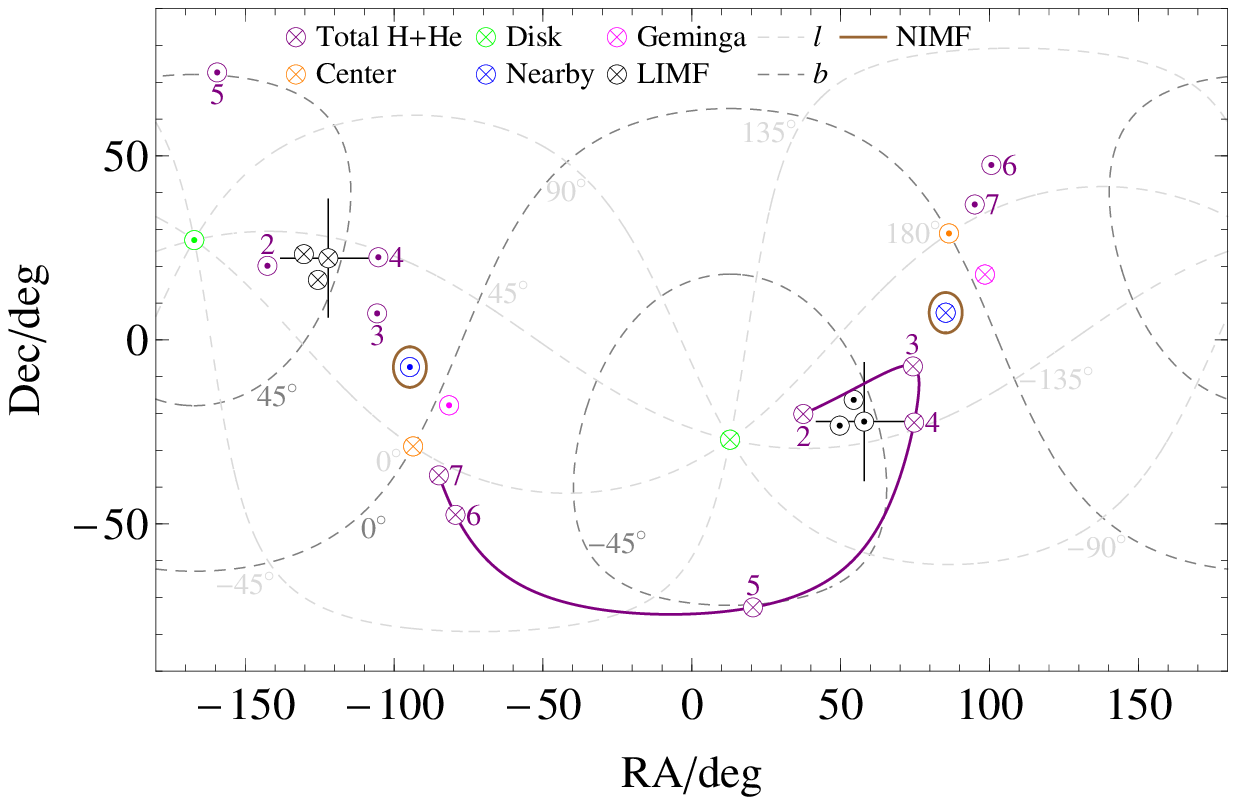}\label{f1d}}
	\caption{CR data fit using the naive approach, where red, orange, green and blue solid lines show contributions of the total, Galactic-center, -disk and nearby-source H component, respectively, and dashed lines with the same colors show He components. Note that the dipole anisotropy component is in the sense of $ A_{ab}f_{ab}dR_b/\sum_{a,b}{f_{ab}dR_b} $, where $ a=\text{c} $, d or n, $ b=\text{H} $ or He. The purple solid line indicates the total anisotropy taking all components into account, whose projection onto the equatorial plane is given by the cyan solid line, which should be compared with the amplitude data. The cross-dot convention is used in the sky map, i.e., the circle-cross and -dot marker indicate the vectorial forward and reverse direction, respectively. Especially, the purple marker corresponds to the (predicted) total anisotropy at the energy of $ E $, with $ \log _{10}\left( E/\text{GeV}\right) $ indicated with the purple number. The current direction of the Geminga pulsar is marked with magenta. The light and thick gray dashed lines show Galactic longitudinal and latitudinal contours, respectively. The brown solid circles show possible directions of the predicted NIMF. The spectral data are from AMS-02 \citep{2015PhRvL.114q1103A,2015PhRvL.115u1101A}, DAMPE \citep{2019SciA....5.3793A,2021PhRvL.126t1102A}, Tibet-AS$ \gamma $ \citep[the SIBYLL+HD model in][]{2006PhLB..632...58T} and KASCADE \citep[the SIBYLL model in][]{2005APh....24....1A}. The anisotropy data are from underground muon detectors \citep[UG-$ \mu $ data in][]{2005ApJ...626L..29A}, ARGO-YBJ \citep{2015ApJ...809...90B,2018ApJ...861...93B}, Tibet-AS$ \gamma $ \citep{2005ApJ...626L..29A,2017ApJ...836..153A}, EAS-TOP \citep{1995ICRC....2..800A,1996ApJ...470..501A,2009ApJ...692L.130A}, IceCube/Top \citep{2010ApJ...718L.194A,2012ApJ...746...33A,2013ApJ...765...55A,2016ApJ...826..220A} and KASCADE-Grande \citep{2019ApJ...870...91A}. The LIMF data are from the IBEX observation \citep{2013ApJ...776...30F,2015ApJ...814..112F,2016ApJ...818L..18Z}.}\label{f1}
\end{figure}

\begin{figure}[b]
	\centering
	\subfigure[Spectra]{\includegraphics[width=0.496\textwidth]{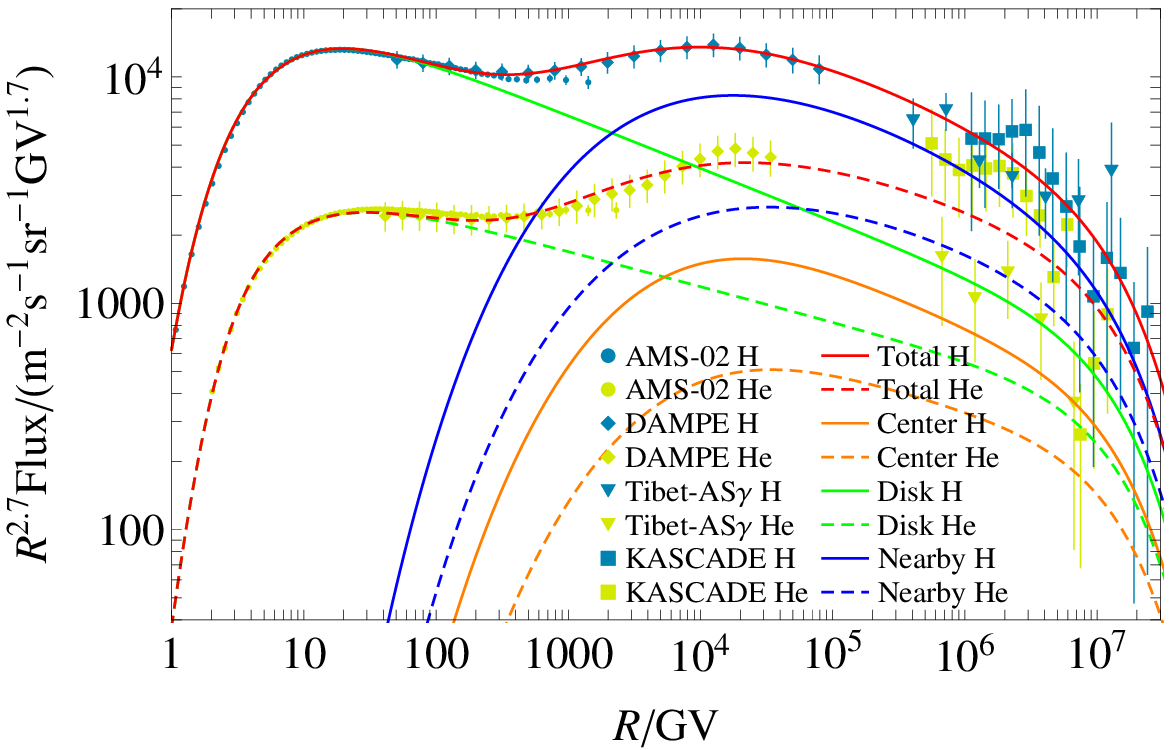}\label{f2a}}
	\subfigure[Dipole Anisotropy Amplitude]{\includegraphics[width=0.496\textwidth]{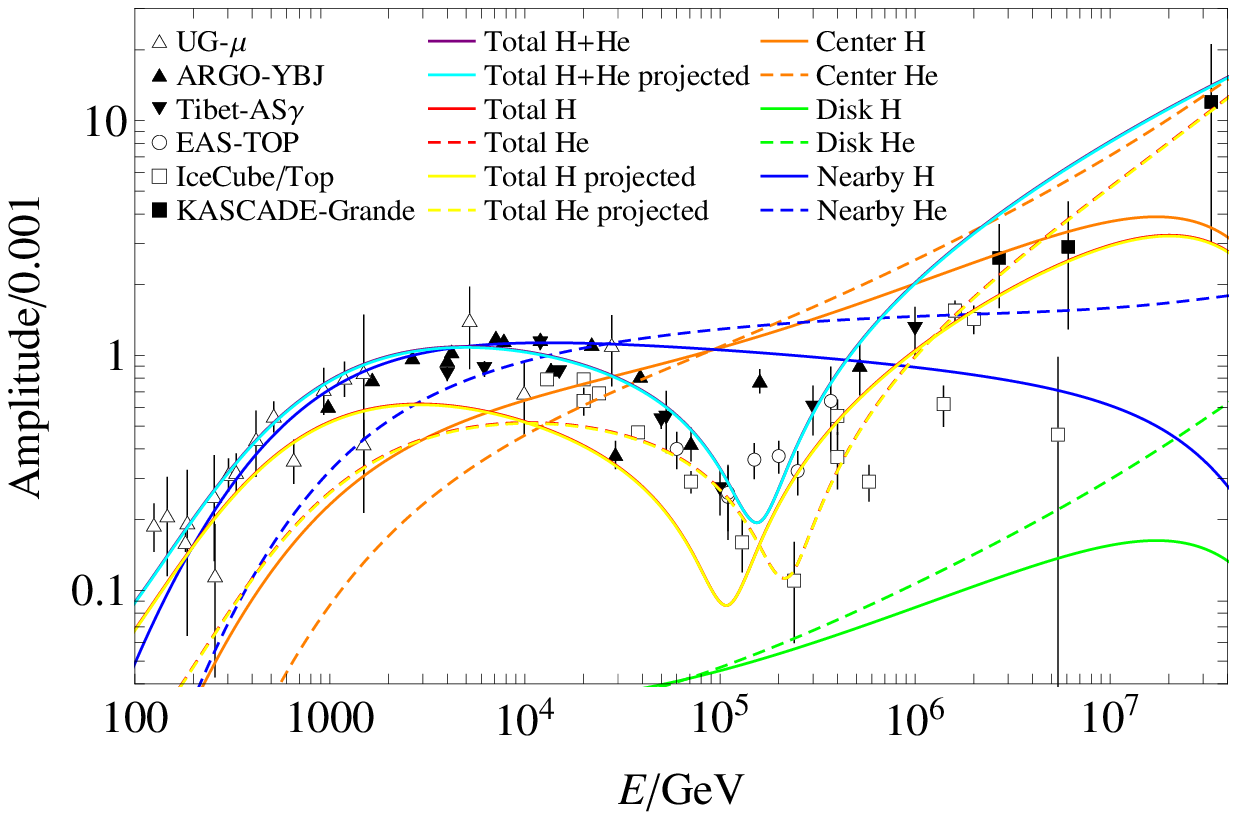}\label{f2b}}
	\subfigure[Dipole Anisotropy Phase]{\includegraphics[width=0.496\textwidth]{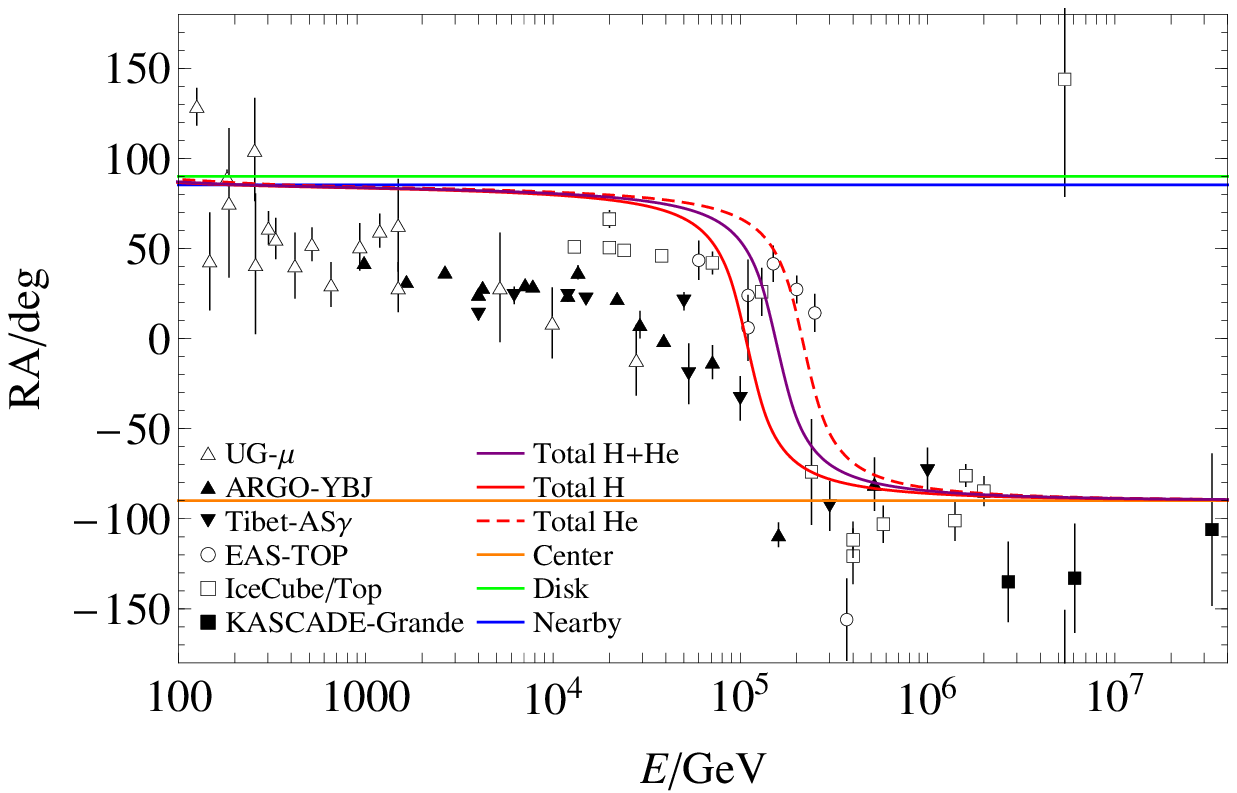}\label{f2c}}
	\subfigure[Sky Map]{\includegraphics[width=0.496\textwidth]{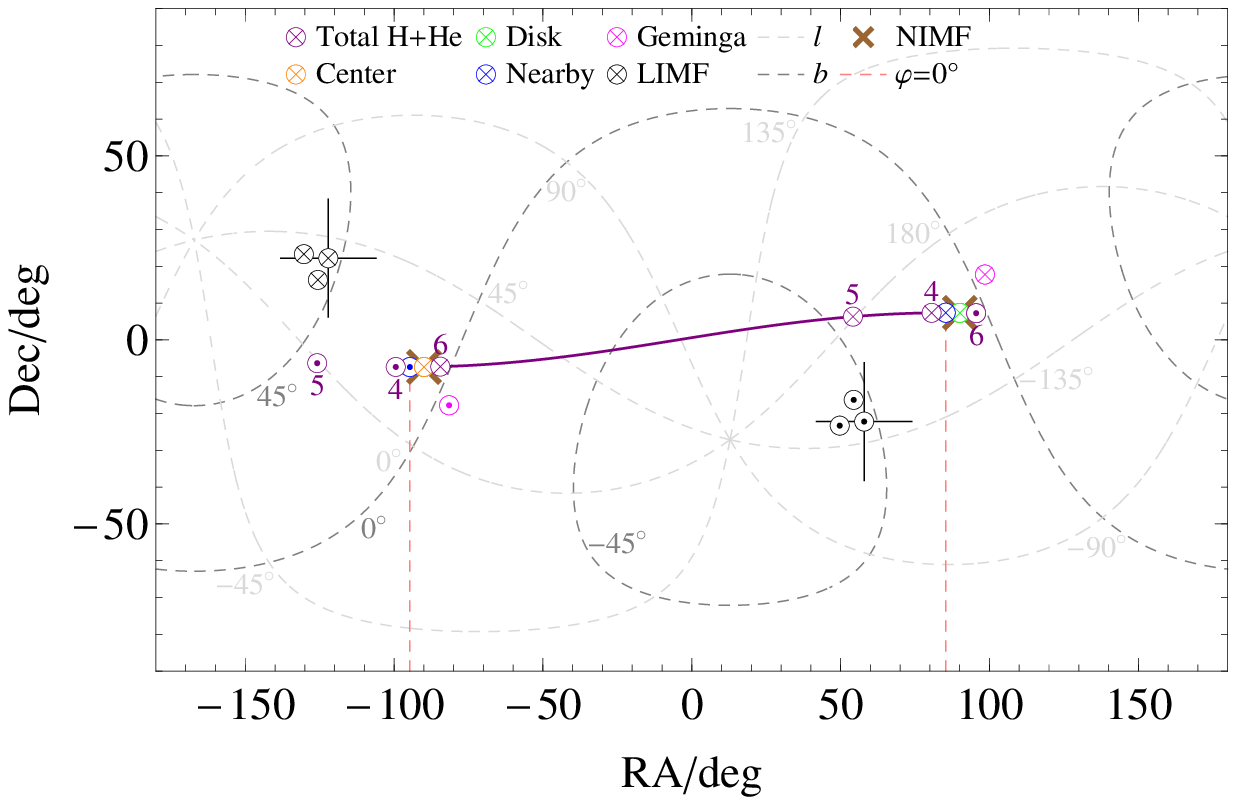}\label{f2d}}
	\caption{CR data fit using the quasi-local approach, where the disk and center anisotropy component after modulations by the NIMF are plotted, and most of the plot markers have been expounded in the caption of Fig.~\ref{f1}. Besides, the brown cross marker shows the unique field-line alignment of the NIMF determined by this approach. We choose $ \varphi =0^{\circ} $ as the pink dashed line, i.e., the $ 0^{\circ} $ longitude in a coordinate system whose pole is the line-of-sight vector toward the nearby source. Note that the sky map is here plotted with the left-handed coordinate.}\label{f2}
\end{figure}

To further reduce the number of fitting parameters, we assume that no source is unique in terms of the acceleration spectral shape, i.e., all the source components have the same H spectral index $ \alpha $, He spectral index, maximum rigidity $ R_\text{m} $, and He-to-H Galactic injection flux ratio $ \chi $. The spectrum of He seems harder than that of H as indicated by many spectral studies, implying, e.g., time-dependent particle acceleration \citep{2017ApJ...844L...3Z}, or more efficient He injection in the shock acceleration process \citep{2012PhRvL.108h1104M}. It is thus assumed the He spectral index to be $ \alpha -0.077 $ according to the sub-TeV data fit \citep{2015PhRvL.115u1101A}. To be consistent with the conventional scenario, we also consider that the nearby SNR can inject a total CR energy of $ 10^{50} $ erg into the ISM, i.e.,
\begin{align}
\int_{\text{GeV}}^{\infty}{\left[ \frac{dR_{\text{H}}}{dE}+\chi \left( E \right) \frac{dR_{\text{He}}}{dE} \right] N\left( E \right) EdE}=10^{50}\text{ erg},
\end{align}
where $ E $ is the particle kinetic energy, $ N $ refers to the injection spectrum of H with $ N\left( \text{GV}\right) \approx 4.7\times 10^{52}\text{ GV}^{-1} $ (for the fitted parameters in Tab.~\ref{t1}). For brevity, we hereafter treat all of $ N $ and $ Q $ as quantities for H components. With these respects, in the naive approach, the spectral fitting parameters are $ \chi \left( \text{GV}\right) $, $ \alpha $, $ R_\text{m} $, $ \phi $, $ \theta $, $ M_\text{A} $, $ \left| z/h \right| $, $ \mu Q_{\text{d}}\left( \text{GV}\right) $, $ Q_{\text{c}}\left( \text{GV}\right) $, $ t_{\text{c}} $, where $ \theta =\arcsin \left( \rho /r_{\text{n}} \right) $ is the angle between the regular magnetic field line and the line-of-sight vector toward the nearby source. In the quasi-local approach Eq.~(\ref{Aql}), the alignment of the field line must uniquely be determined, leading to an additional angular parameter, which can be chosen as the angle $ \varphi $ around the line-of-sight vector (toward the nearby source with $ \varphi =0^{\circ} $ shown in Fig.~\ref{f2d}). The fitting results of the CR spectrum for models with the naive and quasi-local approach are shown with Figs.~\ref{f1a} and \ref{f2a}, respectively.

These spectral results are not independent of the dipole anisotropy fit. To calculate the anisotropy, we should further know directions of the sources. It is said that the current proper motion of Geminga along the celestial sphere is of an RA velocity about 0.14 $ \text{arcsec}/\text{yr} $, and a Dec velocity about 0.11 $ \text{arcsec}/\text{yr} $ \citep{2007Ap&SS.308..225F}. Assuming a uniform proper motion from the time of birth (i.e., 340 kyr ago), we estimate the equatorial coordinate of the birth place of Geminga to be $ \left( \text{RA},\text{Dec} \right) \approx \left( 85.3,7.38 \right) ^{\circ} $, corresponding to a Galactic coordinate $ \left( l,b \right) \approx -\left( 162,12 \right) ^{\circ} $, which is the direction of $ \boldsymbol{A}_{\text{n}} $. By definition, the Galactic latitude of the direction of $ \left< \boldsymbol{A}_{\text{d}} \right> $ is $ -90^{\circ} $, and $ \left< \boldsymbol{A}_{\text{c}} \right> $ is toward the Galactic coordinate $ \left( 0,0 \right) ^{\circ} $.

It should be noted that many ground-based observations only report a mean dipole anisotropy projected onto the equatorial plane due to observational limitations \citep{2017PrPNP..94..184A}. This is the case for the anisotropy data involved in this work, thus we should also use the projection of calculated $ A $ onto the equatorial plane, i.e., $ A\cos \left( \text{Dec of }\boldsymbol{A} \right) $, to compare with the observations. The anisotropy fitting results are shown with Figs.~\ref{f1b}, \ref{f1c}, \ref{f2b} and \ref{f2c}, correspondingly the predicted dipole directions in the sky map are shown in Figs.~\ref{f1d} and \ref{f2d}, and values of all the fitting parameters are given by Tab.~\ref{t1}.
\section{Discussion}
As seen from Figs.~\ref{f1b} and \ref{f2b}, in the naive approach there is a significant projection effect of the dipole anisotropy onto the equatorial plane serving as an important reason for the amplitude dip around 100 TeV, while the projection effect is not important in the quasi-local approach for $ \varphi =90^{\circ} $. These two cases correspond to the large and small Dec of the predicted dipole direction shown in Figs.~\ref{f1d} and \ref{f2d}, respectively. This Dec uncertainty needs to be reduced with 3D dipole data. The dip arises from the competition of the nearby and center component, since the two components are in largely opposite directions, and $ \left< \boldsymbol{A}_{\text{c}} \right> $ (which is proportional to $ D $ in the high-energy limit) increases with $ E $ faster than $ \boldsymbol{A}_{\text{n}} $ (which is energy-independent). The amplitude fit seems to suggest a slightly lower $ \varDelta $ than $ 1/3 $ above 100 TeV. It is also possible that the high-energy anisotropy is reduced by the two-halo diffusion effect of Galactic CRs \citep{2012ApJ...752L..13T}. The existence of heavier nuclei than He may also suppress the increase of the anisotropy above PeV \citep{2019JCAP...12..007Q}.

Figs.~\ref{f1c} and \ref{f2c} show that both these approaches can produce the phase-flip behavior of the dipole anisotropy, but give relatively poor fits to the RA data below 100 TeV. Intuitively, the data seem to suggest that the nearby source is in fact in the RA range about 30--$ 60^{\circ} $, where no source that may account for the spectral bump has been detected. Within the presented framework, a possible solution is further taking into account the modulation of low-energy CR diffusion by the local interstellar magnetic field (LIMF) in the vicinity of the Solar system, as observations show that this field may roughly be aligned with the TeV dipole anisotropy \citep{2016PhRvL.117o1103A}. For PeV CRs, since their gyro-radii are large, and the anisotropy is dominated by the central source, the LIMF may not affect the RA phase significantly.

Nevertheless, the three-component model provides a possible scenario about the CR source configuration and multi-scale diffusion. The parameters in Tab.~\ref{t1} suggest that Galactic CR sources are PeVatrons capable of accelerating particles up to 20 PV with a soft spectrum of $ \alpha =2.6 $ injected into the ISM, and the three components can thus contribute to the all-particle distribution up to 500 PeV (by iron nuclei). The injection spectral shape of the nearby source is not unique compared with that of the background sources. Moreover, the universality of the total energy injected by an SNR, i.e., $ 10^{50} $ erg, can be achieved if $ \mu \sim 5/\pi \text{ kpc}^{-2} $ and each SNR on the Galactic disk has an effective lifetime of 100 kyr. For a disk with a radius of 15 kpc, this requires the explosion rate of Galactic supernovae to be on the order of $ 0.01\text{ yr}^{-1} $. Such a total energy for the nearby source, along with the TeV low-energy cutoff, also requires a small value of $ \theta $ and $ M_{\text{A}} $. If the nearby source is a hypernova with more energy input into CRs, both $ \theta $ and $ M_{\text{A}} $ can have higher values. The $ \varphi =90^{\circ} $ minimizes the RA phase below 100 TeV in Fig.~\ref{f2c}, from which we would suggest a nearby interstellar magnetic field (NIMF, i.e., the regular magnetic field on the Geminga-to-Earth scale presumed in previous sections) directed at the equatorial coordinate about $ -\left( 90,7.4 \right) ^{\circ} $ assuming a minimal deflection from the NIMF to LIMF (see Fig.~\ref{f2d}). This direction also differs from that of the magnetic field in the vicinity of the Geminga PWN expected by the anisotropic diffusion scenario \citep{2019PhRvL.123v1103L}.

The model also indicates that the Galactic center region has been continuously injecting CRs for 18 Myr, with an injection power on the order of $ 10^{53}\text{ erg}/\left( 100\text{ kyr} \right) $, which is comparable with thousands of SNR-like sources. If the power is contributed by an ensemble of SNRs around the center, a higher explosion rate of supernovae than that before 18 Myr, which may correspond to a starburst phase in even earlier years, should be maintained up to now in the center region. The power may also arise from the continuous particle acceleration of Sgr A$ ^* $. In any case, there seems to be an event tens of Myr ago that triggered the continuous CR injection. Even the event is strong, typically CRs from the center burst injection can not well be observed at Earth because of the CR escape and finite diffusion time (see Eq.~(\ref{G}) for $ L<\varrho $). The expected $ t_{\text{c}} $ is also in agreement with the age of Fermi bubbles indicated by the star-formation and hot-accretion wind model \citep{2018ApJ...869L..20M}. Further studies should be based on detailed investigation of the Galactic center region, signal detection of the nearby source radiation, and on more precise $ \gamma $-ray and CR measurements such as LHAASO \citep{2019arXiv190502773B}.
\section*{Acknowledgment}
This work is partially supported by grants from the National Natural Science Foundation of China (Grant No.~11761131007, U1738122, U1931204, U2031111), National Key R\&D Program of China (2018YFA0404203), International Partnership Program of Chinese Academy of Sciences (114332KYSB20170008), and China Scholarship Council (201806340077).
\section*{Data Availability}
All results in this paper are obtained using publicly available data.
\bibliographystyle{aasjournal}
\bibliography{ref}

\begin{thebibliography}{}
\expandafter\ifx\csname natexlab\endcsname\relax\def\natexlab#1{#1}\fi

\bibitem[{{Aartsen} {et~al.}(2013){Aartsen}, {Abbasi}, {Abdou}, {Ackermann},
  {Adams}, {Aguilar}, {Ahlers}, {Altmann}, {Andeen}, {Auffenberg}, {Bai},
  {Baker}, {Barwick}, {Baum}, {Bay}, {Beattie}, {Beatty}, {Bechet}, {Becker
  Tjus}, {Becker}, {Bell}, {Benabderrahmane}, {BenZvi}, {Berdermann},
  {Berghaus}, {Berley}, {Bernardini}, {Bertrand}, {Besson}, {Bindig}, {Bissok},
  {Blaufuss}, {Blumenthal}, {Boersma}, {Bohaichuk}, {Bohm}, {Bose},
  {B{\"o}ser}, {Botner}, {Brayeur}, {Brown}, {Bruijn}, {Brunner}, {Carson},
  {Casey}, {Casier}, {Chirkin}, {Christy}, {Clark}, {Clevermann}, {Cohen},
  {Cowen}, {Cruz Silva}, {Danninger}, {Daughhetee}, {Davis}, {De Clercq}, {De
  Ridder}, {Descamps}, {Desiati}, {de Vries-Uiterweerd}, {DeYoung},
  {D{\'\i}az-V{\'e}lez}, {Dreyer}, {Dumm}, {Dunkman}, {Eagan}, {Eisch},
  {Ellsworth}, {Engdeg{\r{a}}rd}, {Euler}, {Evenson}, {Fadiran}, {Fazely},
  {Fedynitch}, {Feintzeig}, {Feusels}, {Filimonov}, {Finley}, {Fischer-Wasels},
  {Flis}, {Franckowiak}, {Franke}, {Frantzen}, {Fuchs}, {Gaisser}, {Gallagher},
  {Gerhardt}, {Gladstone}, {Gl{\"u}senkamp}, {Goldschmidt}, {Golup}, {Goodman},
  {G{\'o}ra}, {Grant}, {Gross}, {Grullon}, {Gurtner}, {Ha}, {Haj Ismail},
  {Hallgren}, {Halzen}, {Hanson}, {Heereman}, {Heimann}, {Heinen}, {Helbing},
  {Hellauer}, {Hickford}, {Hill}, {Hoffman}, {Hoffmann}, {Homeier}, {Hoshina},
  {Huelsnitz}, {Hulth}, {Hultqvist}, {Hussain}, {Ishihara}, {Jacobi},
  {Jacobsen}, {Japaridze}, {Jlelati}, {Kappes}, {Karg}, {Karle}, {Kiryluk},
  {Kislat}, {Kl{\"a}s}, {Klein}, {K{\"o}hne}, {Kohnen}, {Kolanoski},
  {K{\"o}pke}, {Kopper}, {Kopper}, {Koskinen}, {Kowalski}, {Krasberg}, {Kroll},
  {Kunnen}, {Kurahashi}, {Kuwabara}, {Labare}, {Landsman}, {Larson}, {Lauer},
  {Lesiak-Bzdak}, {L{\"u}nemann}, {Madsen}, {Maruyama}, {Mase}, {Matis},
  {McNally}, {Meagher}, {Merck}, {M{\'e}sz{\'a}ros}, {Meures}, {Miarecki},
  {Middell}, {Milke}, {Miller}, {Mohrmann}, {Montaruli}, {Morse}, {Nahnhauer},
  {Naumann}, {Nowicki}, {Nygren}, {Obertacke}, {Odrowski}, {Olivas}, {Olivo},
  {O'Murchadha}, {Panknin}, {Paul}, {Pepper}, {P{\'e}rez de los Heros},
  {Pieloth}, {Pirk}, {Posselt}, {Price}, {Przybylski}, {R{\"a}del}, {Rawlins},
  {Redl}, {Resconi}, {Rhode}, {Ribordy}, {Richman}, {Riedel}, {Rodrigues},
  {Rothmaier}, {Rott}, {Ruhe}, {Ruzybayev}, {Ryckbosch}, {Saba}, {Salameh},
  {Sander}, {Santander}, {Sarkar}, {Schatto}, {Scheel}, {Scheriau}, {Schmidt},
  {Schmitz}, {Schoenen}, {Sch{\"o}neberg}, {Sch{\"o}nherr}, {Sch{\"o}nwald},
  {Schukraft}, {Schulte}, {Schulz}, {Seckel}, {Seo}, {Sestayo}, {Seunarine},
  {Sheremata}, {Smith}, {Soiron}, {Soldin}, {Spiczak}, {Spiering},
  {Stamatikos}, {Stanev}, {Stasik}, {Stezelberger}, {Stokstad}, {St{\"o}ssl},
  {Strahler}, {Str{\"o}m}, {Sullivan}, {Taavola}, {Taboada}, {Tamburro},
  {Ter-Antonyan}, {Tilav}, {Toale}, {Toscano}, {Usner}, {van der Drift}, {van
  Eijndhoven}, {Van Overloop}, {van Santen}, {Vehring}, {Voge}, {Vraeghe},
  {Walck}, {Waldenmaier}, {Wallraff}, {Walter}, {Wasserman}, {Weaver}, {Wendt},
  {Westerhoff}, {Whitehorn}, {Wiebe}, {Wiebusch}, {Williams}, {Wissing},
  {Wolf}, {Wood}, {Woschnagg}, {Xu}, {Xu}, {Xu}, {Yanez}, {Yodh}, {Yoshida},
  {Zarzhitsky}, {Ziemann}, {Zierke}, {Zilles}, {Zoll}, \& {IceCube
  Collaboration}}]{2013ApJ...765...55A}
{Aartsen}, M.~G., {Abbasi}, R., {Abdou}, Y., {et~al.} 2013, \apj, 765, 55

\bibitem[{{Aartsen} {et~al.}(2016){Aartsen}, {Abraham}, {Ackermann}, {Adams},
  {Aguilar}, {Ahlers}, {Ahrens}, {Altmann}, {Anderson}, {Ansseau}, {Anton},
  {Archinger}, {Arguelles}, {Arlen}, {Auffenberg}, {Bai}, {Barwick}, {Baum},
  {Bay}, {Beatty}, {Becker Tjus}, {Becker}, {Beiser}, {BenZvi}, {Berghaus},
  {Berley}, {Bernardini}, {Bernhard}, {Besson}, {Binder}, {Bindig}, {Bissok},
  {Blaufuss}, {Blumenthal}, {Boersma}, {Bohm}, {B{\"o}rner}, {Bos}, {Bose},
  {B{\"o}ser}, {Botner}, {Braun}, {Brayeur}, {Bretz}, {Buzinsky}, {Casey},
  {Casier}, {Cheung}, {Chirkin}, {Christov}, {Clark}, {Classen}, {Coenders},
  {Collin}, {Conrad}, {Cowen}, {Cruz Silva}, {Daughhetee}, {Davis}, {Day}, {de
  Andr{\'e}}, {De Clercq}, {del Pino Rosendo}, {Dembinski}, {De Ridder},
  {Desiati}, {de Vries}, {de Wasseige}, {de With}, {DeYoung},
  {D{\'\i}az-V{\'e}lez}, {di Lorenzo}, {Dujmovic}, {Dumm}, {Dunkman},
  {Eberhardt}, {Ehrhardt}, {Eichmann}, {Euler}, {Evenson}, {Fahey}, {Fazely},
  {Feintzeig}, {Felde}, {Filimonov}, {Finley}, {Flis}, {F{\"o}sig}, {Fuchs},
  {Gaisser}, {Gaior}, {Gallagher}, {Gerhardt}, {Ghorbani}, {Gier}, {Gladstone},
  {Glagla}, {Gl{\"u}senkamp}, {Goldschmidt}, {Golup}, {Gonzalez}, {G{\'o}ra},
  {Grant}, {Griffith}, {Ha}, {Haack}, {Haj Ismail}, {Hallgren}, {Halzen},
  {Hansen}, {Hansmann}, {Hansmann}, {Hanson}, {Hebecker}, {Heereman},
  {Helbing}, {Hellauer}, {Hickford}, {Hignight}, {Hill}, {Hoffman}, {Hoffmann},
  {Holzapfel}, {Homeier}, {Hoshina}, {Huang}, {Huber}, {Huelsnitz}, {Hulth},
  {Hultqvist}, {In}, {Ishihara}, {Jacobi}, {Japaridze}, {Jeong}, {Jero},
  {Jones}, {Jurkovic}, {Kappes}, {Karg}, {Karle}, {Katz}, {Kauer}, {Keivani},
  {Kelley}, {Kemp}, {Kheirandish}, {Kim}, {Kintscher}, {Kiryluk}, {Klein},
  {Kohnen}, {Koirala}, {Kolanoski}, {Konietz}, {K{\"o}pke}, {Kopper}, {Kopper},
  {Koskinen}, {Kowalski}, {Krings}, {Kroll}, {Kroll}, {Kr{\"u}ckl}, {Kunnen},
  {Kunwar}, {Kurahashi}, {Kuwabara}, {Labare}, {Lanfranchi}, {Larson},
  {Lennarz}, {Lesiak-Bzdak}, {Leuermann}, {Leuner}, {Lu}, {L{\"u}nemann},
  {Madsen}, {Maggi}, {Mahn}, {Mandelartz}, {Maruyama}, {Mase}, {Matis},
  {Maunu}, {McNally}, {Meagher}, {Medici}, {Meier}, {Meli}, {Menne}, {Merino},
  {Meures}, {Miarecki}, {Middell}, {Mohrmann}, {Montaruli}, {Morse},
  {Nahnhauer}, {Naumann}, {Neer}, {Niederhausen}, {Nowicki}, {Nygren},
  {Obertacke Pollmann}, {Olivas}, {Omairat}, {O'Murchadha}, {Palczewski},
  {Pandya}, {Pankova}, {Paul}, {Pepper}, {P{\'e}rez de los Heros}, {Pfendner},
  {Pieloth}, {Pinat}, {Posselt}, {Price}, {Przybylski}, {Quinnan}, {Raab},
  {R{\"a}del}, {Rameez}, {Rawlins}, {Reimann}, {Relich}, {Resconi}, {Rhode},
  {Richman}, {Richter}, {Riedel}, {Robertson}, {Rongen}, {Rott}, {Ruhe},
  {Ryckbosch}, {Sabbatini}, {Sander}, {Sandrock}, {Sandroos}, {Sarkar},
  {Schatto}, {Schimp}, {Schlunder}, {Schmidt}, {Schoenen}, {Sch{\"o}neberg},
  {Sch{\"o}nwald}, {Schumacher}, {Seckel}, {Seunarine}, {Soldin}, {Song},
  {Spiczak}, {Spiering}, {Stahlberg}, {Stamatikos}, {Stanev}, {Stasik},
  {Steuer}, {Stezelberger}, {Stokstad}, {St{\"o}ssl}, {Str{\"o}m},
  {Strotjohann}, {Sullivan}, {Sutherland}, {Taavola}, {Taboada}, {Tatar},
  {Ter-Antonyan}, {Terliuk}, {Te{\v{s}}i{\'c}}, {Tilav}, {Toale}, {Tobin},
  {Toscano}, {Tosi}, {Tselengidou}, {Turcati}, {Unger}, {Usner}, {Vallecorsa},
  {Vandenbroucke}, {van Eijndhoven}, {Vanheule}, {van Santen}, {Veenkamp},
  {Vehring}, {Voge}, {Vraeghe}, {Walck}, {Wallace}, {Wallraff}, {Wandkowsky},
  {Weaver}, {Wendt}, {Westerhoff}, {Whelan}, {Wiebe}, {Wiebusch}, {Wille},
  {Williams}, {Wills}, {Wissing}, {Wolf}, {Wood}, {Woschnagg}, {Xu}, {Xu},
  {Xu}, {Yanez}, {Yodh}, {Yoshida}, {Zoll}, \& {IceCube
  Collaboration}}]{2016ApJ...826..220A}
{Aartsen}, M.~G., {Abraham}, K., {Ackermann}, M., {et~al.} 2016, \apj, 826, 220

\bibitem[{{Abbasi} {et~al.}(2010){Abbasi}, {Abdou}, {Abu-Zayyad}, {Adams},
  {Aguilar}, {Ahlers}, {Andeen}, {Auffenberg}, {Bai}, {Baker}, {Barwick},
  {Bay}, {Bazo Alba}, {Beattie}, {Beatty}, {Bechet}, {Becker}, {Becker},
  {Benabderrahmane}, {BenZvi}, {Berdermann}, {Berghaus}, {Berley},
  {Bernardini}, {Bertrand}, {Besson}, {Bissok}, {Blaufuss}, {Boersma}, {Bohm},
  {B{\"o}ser}, {Botner}, {Bradley}, {Braun}, {Buitink}, {Carson}, {Chirkin},
  {Christy}, {Clem}, {Clevermann}, {Cohen}, {Colnard}, {Cowen}, {D'Agostino},
  {Danninger}, {Davis}, {De Clercq}, {Demir{\"o}rs}, {Depaepe}, {Descamps},
  {Desiati}, {de Vries-Uiterweerd}, {DeYoung}, {D{\'\i}az-V{\'e}lez},
  {Dierckxsens}, {Dreyer}, {Dumm}, {Duvoort}, {Ehrlich}, {Eisch}, {Ellsworth},
  {Engdeg{\r{a}}rd}, {Euler}, {Evenson}, {Fadiran}, {Fazely}, {Feusels},
  {Filimonov}, {Finley}, {Foerster}, {Fox}, {Franckowiak}, {Franke}, {Gaisser},
  {Gallagher}, {Geisler}, {Gerhardt}, {Gladstone}, {Gl{\"u}senkamp},
  {Goldschmidt}, {Goodman}, {Grant}, {Griesel}, {Gro{\ss}}, {Grullon},
  {Gurtner}, {Ha}, {Hallgren}, {Halzen}, {Han}, {Hanson}, {Helbing}, {Herquet},
  {Hickford}, {Hill}, {Hoffman}, {Homeier}, {Hoshina}, {Hubert}, {Huelsnitz},
  {H{\"u}l{\ss}}, {Hulth}, {Hultqvist}, {Hussain}, {Ishihara}, {Jacobsen},
  {Japaridze}, {Johansson}, {Joseph}, {Kampert}, {Karg}, {Karle}, {Kelley},
  {Kemming}, {Kenny}, {Kiryluk}, {Kislat}, {Klein}, {Knops}, {K{\"o}hne},
  {Kohnen}, {Kolanoski}, {K{\"o}pke}, {Koskinen}, {Kowalski}, {Kowarik},
  {Krasberg}, {Krings}, {Kroll}, {Kuehn}, {Kuwabara}, {Labare}, {Lafebre},
  {Laihem}, {Landsman}, {Lauer}, {Lehmann}, {Lennarz}, {L{\"u}nemann},
  {Madsen}, {Majumdar}, {Marotta}, {Maruyama}, {Mase}, {Matis}, {Matusik},
  {Meagher}, {Merck}, {M{\'e}sz{\'a}ros}, {Meures}, {Middell}, {Milke},
  {Miller}, {Montaruli}, {Morse}, {Movit}, {Nahnhauer}, {Nam}, {Naumann},
  {Nie{\ss}en}, {Nygren}, {Odrowski}, {Olivas}, {Olivo}, {O'Murchadha}, {Ono},
  {Panknin}, {Paul}, {P{\'e}rez de los Heros}, {Petrovic}, {Piegsa}, {Pieloth},
  {Porrata}, {Posselt}, {Price}, {Prikockis}, {Przybylski}, {Rawlins}, {Redl},
  {Resconi}, {Rhode}, {Ribordy}, {Rizzo}, {Rodrigues}, {Roth}, {Rothmaier},
  {Rott}, {Roucelle}, {Ruhe}, {Rutledge}, {Ruzybayev}, {Ryckbosch}, {Sander},
  {Santander}, {Sarkar}, {Schatto}, {Schlenstedt}, {Schmidt}, {Schukraft},
  {Schultes}, {Schulz}, {Schunck}, {Seckel}, {Semburg}, {Seo}, {Sestayo},
  {Seunarine}, {Silvestri}, {Slipak}, {Spiczak}, {Spiering}, {Stamatikos},
  {Stanev}, {Stephens}, {Stezelberger}, {Stokstad}, {Stoyanov}, {Strahler},
  {Straszheim}, {Sullivan}, {Swillens}, {Taavola}, {Taboada}, {Tamburro},
  {Tarasova}, {Tepe}, {Ter-Antonyan}, {Tilav}, {Toale}, {Toscano}, {Tosi},
  {Tur{\v{c}}an}, {van Eijndhoven}, {Vandenbroucke}, {Van Overloop}, {van
  Santen}, {Voge}, {Voigt}, {Walck}, {Waldenmaier}, {Wallraff}, {Walter},
  {Weaver}, {Wendt}, {Westerhoff}, {Whitehorn}, {Wiebe}, {Wiebusch},
  {Wikstr{\"o}m}, {Williams}, {Wischnewski}, {Wissing}, {Wolf}, {Woschnagg},
  {Xu}, {Xu}, {Yodh}, {Yoshida}, {Zarzhitsky}, \& {IceCube
  Collaboration}}]{2010ApJ...718L.194A}
{Abbasi}, R., {Abdou}, Y., {Abu-Zayyad}, T., {et~al.} 2010, \apjl, 718, L194

\bibitem[{{Abbasi} {et~al.}(2012){Abbasi}, {Abdou}, {Abu-Zayyad}, {Ackermann},
  {Adams}, {Aguilar}, {Ahlers}, {Allen}, {Altmann}, {Andeen}, {Auffenberg},
  {Bai}, {Baker}, {Barwick}, {Bay}, {Bazo Alba}, {Beattie}, {Beatty}, {Bechet},
  {Becker}, {Becker}, {Benabderrahmane}, {BenZvi}, {Berdermann}, {Berghaus},
  {Berley}, {Bernardini}, {Bertrand}, {Besson}, {Bindig}, {Bissok}, {Blaufuss},
  {Blumenthal}, {Boersma}, {Bohm}, {Bose}, {B{\"o}ser}, {Botner}, {Brown},
  {Buitink}, {Caballero-Mora}, {Carson}, {Chirkin}, {Christy}, {Clevermann},
  {Cohen}, {Colnard}, {Cowen}, {Cruz Silva}, {D'Agostino}, {Danninger},
  {Daughhetee}, {Davis}, {De Clercq}, {Degner}, {Demir{\"o}rs}, {Descamps},
  {Desiati}, {de Vries-Uiterweerd}, {DeYoung}, {D{\'\i}az-V{\'e}lez},
  {Dierckxsens}, {Dreyer}, {Dumm}, {Dunkman}, {Eisch}, {Ellsworth},
  {Engdeg{\r{a}}rd}, {Euler}, {Evenson}, {Fadiran}, {Fazely}, {Fedynitch},
  {Feintzeig}, {Feusels}, {Filimonov}, {Finley}, {Fischer-Wasels}, {Fox},
  {Franckowiak}, {Franke}, {Gaisser}, {Gallagher}, {Gerhardt}, {Gladstone},
  {Gl{\"u}senkamp}, {Goldschmidt}, {Goodman}, {G{\'o}ra}, {Grant}, {Griesel},
  {Gro{\ss}}, {Grullon}, {Gurtner}, {Ha}, {Haj Ismail}, {Hallgren}, {Halzen},
  {Han}, {Hanson}, {Heinen}, {Helbing}, {Hellauer}, {Hickford}, {Hill},
  {Hoffman}, {Hoffmann}, {Homeier}, {Hoshina}, {Huelsnitz}, {H{\"u}l{\ss}},
  {Hulth}, {Hultqvist}, {Hussain}, {Ishihara}, {Jacobi}, {Jacobsen},
  {Japaridze}, {Johansson}, {Kampert}, {Kappes}, {Karg}, {Karle}, {Kenny},
  {Kiryluk}, {Kislat}, {Klein}, {K{\"o}hne}, {Kohnen}, {Kolanoski},
  {K{\"o}pke}, {Kopper}, {Koskinen}, {Kowalski}, {Kowarik}, {Krasberg},
  {Kroll}, {Kurahashi}, {Kuwabara}, {Labare}, {Laihem}, {Landsman}, {Larson},
  {Lauer}, {L{\"u}nemann}, {Madsen}, {Marotta}, {Maruyama}, {Mase}, {Matis},
  {Meagher}, {Merck}, {M{\'e}sz{\'a}ros}, {Meures}, {Miarecki}, {Middell},
  {Milke}, {Miller}, {Montaruli}, {Morse}, {Movit}, {Nahnhauer}, {Nam},
  {Naumann}, {Nygren}, {Odrowski}, {Olivas}, {Olivo}, {O'Murchadha}, {Panknin},
  {Paul}, {P{\'e}rez de los Heros}, {Petrovic}, {Piegsa}, {Pieloth}, {Porrata},
  {Posselt}, {Price}, {Price}, {Przybylski}, {Rawlins}, {Redl}, {Resconi},
  {Rhode}, {Ribordy}, {Richman}, {Rodrigues}, {Rothmaier}, {Rott}, {Ruhe},
  {Rutledge}, {Ruzybayev}, {Ryckbosch}, {Sander}, {Santander}, {Sarkar},
  {Schatto}, {Schmidt}, {Sch{\"o}nwald}, {Schukraft}, {Schultes}, {Schulz},
  {Schunck}, {Seckel}, {Semburg}, {Seo}, {Sestayo}, {Seunarine}, {Silvestri},
  {Spiczak}, {Spiering}, {Stamatikos}, {Stanev}, {Stezelberger}, {Stokstad},
  {St{\"o}{\ss}l}, {Strahler}, {Str{\"o}m}, {St{\"u}er}, {Sullivan},
  {Swillens}, {Taavola}, {Taboada}, {Tamburro}, {Tepe}, {Ter-Antonyan},
  {Tilav}, {Toale}, {Toscano}, {Tosi}, {van Eijndhoven}, {Vandenbroucke}, {Van
  Overloop}, {van Santen}, {Vehring}, {Voge}, {Walck}, {Waldenmaier},
  {Wallraff}, {Walter}, {Weaver}, {Wendt}, {Westerhoff}, {Whitehorn}, {Wiebe},
  {Wiebusch}, {Williams}, {Wischnewski}, {Wissing}, {Wolf}, {Wood},
  {Woschnagg}, {Xu}, {Xu}, {Xu}, {Yanez}, {Yodh}, {Yoshida}, {Zarzhitsky},
  {Zoll}, \& {IceCube Collaboration}}]{2012ApJ...746...33A}
---. 2012, \apj, 746, 33

\bibitem[{{Abeysekara} {et~al.}(2017){Abeysekara}, {Albert}, {Alfaro},
  {Alvarez}, {{\'A}lvarez}, {Arceo}, {Arteaga-Vel{\'a}zquez}, {Avila Rojas},
  {Ayala Solares}, {Barber}, {Bautista-Elivar}, {Becerril}, {Belmont-Moreno},
  {BenZvi}, {Berley}, {Bernal}, {Braun}, {Brisbois}, {Caballero-Mora},
  {Capistr{\'a}n}, {Carrami{\~n}ana}, {Casanova}, {Castillo}, {Cotti},
  {Cotzomi}, {Couti{\~n}o de Le{\'o}n}, {De Le{\'o}n}, {De la Fuente},
  {Dingus}, {DuVernois}, {D{\'\i}az-V{\'e}lez}, {Ellsworth}, {Engel},
  {Enr{\'\i}quez-Rivera}, {Fiorino}, {Fraija}, {Garc{\'\i}a-Gonz{\'a}lez},
  {Garfias}, {Gerhardt}, {Gonz{\'a}lez Mu{\~n}oz}, {Gonz{\'a}lez}, {Goodman},
  {Hampel-Arias}, {Harding}, {Hern{\'a}ndez}, {Hern{\'a}ndez-Almada}, {Hinton},
  {Hona}, {Hui}, {H{\"u}ntemeyer}, {Iriarte}, {Jardin-Blicq}, {Joshi},
  {Kaufmann}, {Kieda}, {Lara}, {Lauer}, {Lee}, {Lennarz}, {Vargas},
  {Linnemann}, {Longinotti}, {Luis Raya}, {Luna-Garc{\'\i}a}, {L{\'o}pez-Coto},
  {Malone}, {Marinelli}, {Martinez}, {Martinez-Castellanos},
  {Mart{\'\i}nez-Castro}, {Mart{\'\i}nez-Huerta}, {Matthews},
  {Miranda-Romagnoli}, {Moreno}, {Mostaf{\'a}}, {Nellen}, {Newbold}, {Nisa},
  {Noriega-Papaqui}, {Pelayo}, {Pretz}, {P{\'e}rez-P{\'e}rez}, {Ren}, {Rho},
  {Rivi{\`e}re}, {Rosa-Gonz{\'a}lez}, {Rosenberg}, {Ruiz-Velasco}, {Salazar},
  {Salesa Greus}, {Sandoval}, {Schneider}, {Schoorlemmer}, {Sinnis}, {Smith},
  {Springer}, {Surajbali}, {Taboada}, {Tibolla}, {Tollefson}, {Torres},
  {Ukwatta}, {Vianello}, {Weisgarber}, {Westerhoff}, {Wisher}, {Wood},
  {Yapici}, {Yodh}, {Younk}, {Zepeda}, {Zhou}, {Guo}, {Hahn}, {Li}, \&
  {Zhang}}]{2017Sci...358..911A}
{Abeysekara}, A.~U., {Albert}, A., {Alfaro}, R., {et~al.} 2017, Science, 358,
  911

\bibitem[{{Aglietta} {et~al.}(1995){Aglietta}, {Alessandro}, {Antonioli},
  {Arneodo}, {Bergamasco}, {Bertaina}, {Bosio}, {Castellina}, {Castagnoli},
  {Chaivasa}, {Cini}, {D' Ettorre Piazzoli}, {Di Sciascio}, {Fulgione},
  {Galeotti}, {Ghia}, {Iacovacci}, {Mannocchi}, {Melagrana}, {Mengotti Silva},
  {Morello}, {Navarra}, {Riccati}, {Saavedra}, {Trinchero}, {Vallania}, \&
  {Vernetto}}]{1995ICRC....2..800A}
{Aglietta}, M., {Alessandro}, B., {Antonioli}, P., {et~al.} 1995, in
  International Cosmic Ray Conference, Vol.~2, International Cosmic Ray
  Conference, 800

\bibitem[{{Aglietta} {et~al.}(1996){Aglietta}, {Alessandro}, {Antonioli},
  {Arneodo}, {Bergamasco}, {Bertaina}, {Bosio}, {Castellina}, {Castagnoli},
  {Chiavassa}, {Cini Castagnoli}, {D'Ettorre Piazzoli}, {di Sciascio},
  {Fulgione}, {Galeotti}, {Ghia}, {Iacovacci}, {Mannocchi}, {Melagrana},
  {Mengotti Silva}, {Morello}, {Navarra}, {Riccati}, {Saavedra}, {Trinchero},
  {Vallania}, {Vernetto}, \& {EAS-Top Collaboration}}]{1996ApJ...470..501A}
{Aglietta}, M., {Alessandro}, B., {Antonioli}, P., {et~al.} 1996, \apj, 470,
  501

\bibitem[{{Aglietta} {et~al.}(2009){Aglietta}, {Alekseenko}, {Alessandro},
  {Antonioli}, {Arneodo}, {Bergamasco}, {Bertaina}, {Bonino}, {Castellina},
  {Chiavassa}, {D'Ettorre Piazzoli}, {Di Sciascio}, {Fulgione}, {Galeotti},
  {Ghia}, {Iacovacci}, {Mannocchi}, {Morello}, {Navarra}, {Saavedra},
  {Stamerra}, {Trinchero}, {Valchierotti}, {Vallania}, {Vernetto}, \&
  {Vigorito}}]{2009ApJ...692L.130A}
{Aglietta}, M., {Alekseenko}, V.~V., {Alessandro}, B., {et~al.} 2009, \apjl,
  692, L130

\bibitem[{{Aguilar} {et~al.}(2015{\natexlab{a}}){Aguilar}, {Aisa}, {Alpat},
  {Alvino}, {Ambrosi}, {Andeen}, {Arruda}, {Attig}, {Azzarello}, {Bachlechner},
  {Barao}, {Barrau}, {Barrin}, {Bartoloni}, {Basara}, {Battarbee}, {Battiston},
  {Bazo}, {Becker}, {Behlmann}, {Beischer}, {Berdugo}, {Bertucci}, {Bindi},
  {Bizzaglia}, {Bizzarri}, {Boella}, {de Boer}, {Bollweg}, {Bonnivard},
  {Borgia}, {Borsini}, {Boschini}, {Bourquin}, {Burger}, {Cadoux}, {Cai},
  {Capell}, {Caroff}, {Casaus}, {Castellini}, {Cernuda}, {Cerreta}, {Cervelli},
  {Chae}, {Chang}, {Chen}, {Chen}, {Chen}, {Chen}, {Cheng}, {Chou},
  {Choumilov}, {Choutko}, {Chung}, {Clark}, {Clavero}, {Coignet}, {Consolandi},
  {Contin}, {Corti}, {Gil}, {Coste}, {Creus}, {Crispoltoni}, {Cui}, {Dai},
  {Delgado}, {Della Torre}, {Demirk{\"o}z}, {Derome}, {Di Falco}, {Di Masso},
  {Dimiccoli}, {D{\'\i}az}, {von Doetinchem}, {Donnini}, {Duranti}, {D'Urso},
  {Egorov}, {Eline}, {Eppling}, {Eronen}, {Fan}, {Farnesini}, {Feng},
  {Fiandrini}, {Fiasson}, {Finch}, {Fisher}, {Formato}, {Galaktionov},
  {Gallucci}, {Garc{\'\i}a}, {Garc{\'\i}a-L{\'o}pez}, {Gargiulo}, {Gast},
  {Gebauer}, {Gervasi}, {Ghelfi}, {Giovacchini}, {Goglov}, {Gong}, {Goy},
  {Grabski}, {Grandi}, {Graziani}, {Guandalini}, {Guerri}, {Guo}, {Haas},
  {Habiby}, {Haino}, {Han}, {He}, {Heil}, {Hoffman}, {Hsieh}, {Huang}, {Huh},
  {Incagli}, {Ionica}, {Jang}, {Jinchi}, {Kanishev}, {Kim}, {Kim}, {Kirn},
  {Korkmaz}, {Kossakowski}, {Kounina}, {Kounine}, {Koutsenko}, {Krafczyk}, {La
  Vacca}, {Laudi}, {Laurenti}, {Lazzizzera}, {Lebedev}, {Lee}, {Lee}, {Leluc},
  {Li}, {Li}, {Li}, {Li}, {Li}, {Li}, {Li}, {Li}, {Li}, {Li}, {Lim}, {Lin},
  {Lipari}, {Lippert}, {Liu}, {Liu}, {Liu}, {Lolli}, {Lomtadze}, {Lu}, {Lu},
  {Lu}, {Luebelsmeyer}, {Luo}, {Luo}, {Lv}, {Majka}, {Ma{\~n}{\'a}},
  {Mar{\'\i}n}, {Martin}, {Mart{\'\i}nez}, {Masi}, {Maurin}, {Menchaca-Rocha},
  {Meng}, {Mo}, {Morescalchi}, {Mott}, {M{\"u}ller}, {Nelson}, {Ni}, {Nikonov},
  {Nozzoli}, {Nunes}, {Obermeier}, {Oliva}, {Orcinha}, {Palmonari},
  {Palomares}, {Paniccia}, {Papi}, {Pauluzzi}, {Pedreschi}, {Pensotti},
  {Pereira}, {Picot-Clemente}, {Pilo}, {Piluso}, {Pizzolotto}, {Plyaskin},
  {Pohl}, {Poireau}, {Putze}, {Quadrani}, {Qi}, {Qin}, {Qu}, {R{\"a}ih{\"a}},
  {Rancoita}, {Rapin}, {Ricol}, {Rodr{\'\i}guez}, {Rosier-Lees}, {Rozhkov},
  {Rozza}, {Sagdeev}, {Sandweiss}, {Saouter}, {Schael}, {Schmidt}, {von
  Dratzig}, {Schwering}, {Scolieri}, {Seo}, {Shan}, {Shan}, {Shi}, {Shi},
  {Shi}, {Siedenburg}, {Son}, {Song}, {Spada}, {Spinella}, {Sun}, {Sun},
  {Tacconi}, {Tang}, {Tang}, {Tang}, {Tao}, {Tescaro}, {Ting}, {Ting},
  {Tomassetti}, {Torsti}, {T{\"u}rko{\v{g}}lu}, {Urban}, {Vagelli}, {Valente},
  {Vannini}, {Valtonen}, {Vaurynovich}, {Vecchi}, {Velasco}, {Vialle},
  {Vitale}, {Vitillo}, {Wang}, {Wang}, {Wang}, {Wang}, {Wang}, {Wang}, {Weng},
  {Whitman}, {Wienkenh{\"o}ver}, {Willenbrock}, {Wu}, {Wu}, {Xia}, {Xie},
  {Xie}, {Xiong}, {Xu}, {Xu}, {Yan}, {Yang}, {Yang}, {Yang}, {Ye}, {Yi}, {Yu},
  {Yu}, {Zeissler}, {Zhang}, {Zhang}, {Zhang}, {Zhang}, {Zhang}, {Zhang},
  {Zhang}, {Zheng}, {Zhuang}, {Zhukov}, {Zichichi}, {Zimmermann}, {Zuccon}, \&
  {AMS Collaboration}}]{2015PhRvL.115u1101A}
{Aguilar}, M., {Aisa}, D., {Alpat}, B., {et~al.} 2015{\natexlab{a}}, \prl, 115,
  211101

\bibitem[{{Aguilar} {et~al.}(2015{\natexlab{b}}){Aguilar}, {Aisa}, {Alpat},
  {Alvino}, {Ambrosi}, {Andeen}, {Arruda}, {Attig}, {Azzarello}, {Bachlechner},
  {Barao}, {Barrau}, {Barrin}, {Bartoloni}, {Basara}, {Battarbee}, {Battiston},
  {Bazo}, {Becker}, {Behlmann}, {Beischer}, {Berdugo}, {Bertucci},
  {Bigongiari}, {Bindi}, {Bizzaglia}, {Bizzarri}, {Boella}, {de Boer},
  {Bollweg}, {Bonnivard}, {Borgia}, {Borsini}, {Boschini}, {Bourquin},
  {Burger}, {Cadoux}, {Cai}, {Capell}, {Caroff}, {Casaus}, {Cascioli},
  {Castellini}, {Cernuda}, {Cerreta}, {Cervelli}, {Chae}, {Chang}, {Chen},
  {Chen}, {Cheng}, {Chen}, {Cheng}, {Chou}, {Choumilov}, {Choutko}, {Chung},
  {Clark}, {Clavero}, {Coignet}, {Consolandi}, {Contin}, {Corti}, {Gil},
  {Coste}, {Creus}, {Crispoltoni}, {Cui}, {Dai}, {Delgado}, {Della Torre},
  {Demirk{\"o}z}, {Derome}, {Di Falco}, {Di Masso}, {Dimiccoli}, {D{\'\i}az},
  {von Doetinchem}, {Donnini}, {Du}, {Duranti}, {D'Urso}, {Eline}, {Eppling},
  {Eronen}, {Fan}, {Farnesini}, {Feng}, {Fiandrini}, {Fiasson}, {Finch},
  {Fisher}, {Galaktionov}, {Gallucci}, {Garc{\'\i}a}, {Garc{\'\i}a-L{\'o}pez},
  {Gargiulo}, {Gast}, {Gebauer}, {Gervasi}, {Ghelfi}, {Gillard}, {Giovacchini},
  {Goglov}, {Gong}, {Goy}, {Grabski}, {Grandi}, {Graziani}, {Guandalini},
  {Guerri}, {Guo}, {Haas}, {Habiby}, {Haino}, {Han}, {He}, {Heil}, {Hoffman},
  {Hsieh}, {Huang}, {Huh}, {Incagli}, {Ionica}, {Jang}, {Jinchi}, {Kanishev},
  {Kim}, {Kim}, {Kirn}, {Kossakowski}, {Kounina}, {Kounine}, {Koutsenko},
  {Krafczyk}, {La Vacca}, {Laudi}, {Laurenti}, {Lazzizzera}, {Lebedev}, {Lee},
  {Lee}, {Leluc}, {Levi}, {Li}, {Li}, {Li}, {Li}, {Li}, {Li}, {Li}, {Li}, {Li},
  {Lim}, {Lin}, {Lipari}, {Lippert}, {Liu}, {Liu}, {Lolli}, {Lomtadze}, {Lu},
  {Lu}, {Lu}, {Luebelsmeyer}, {Luo}, {Lv}, {Majka}, {Ma{\~n}{\'a}},
  {Mar{\'\i}n}, {Martin}, {Mart{\'\i}nez}, {Masi}, {Maurin}, {Menchaca-Rocha},
  {Meng}, {Mo}, {Morescalchi}, {Mott}, {M{\"u}ller}, {Ni}, {Nikonov},
  {Nozzoli}, {Nunes}, {Obermeier}, {Oliva}, {Orcinha}, {Palmonari},
  {Palomares}, {Paniccia}, {Papi}, {Pauluzzi}, {Pedreschi}, {Pensotti},
  {Pereira}, {Picot-Clemente}, {Pilo}, {Piluso}, {Pizzolotto}, {Plyaskin},
  {Pohl}, {Poireau}, {Postaci}, {Putze}, {Quadrani}, {Qi}, {Qin}, {Qu},
  {R{\"a}ih{\"a}}, {Rancoita}, {Rapin}, {Ricol}, {Rodr{\'\i}guez},
  {Rosier-Lees}, {Rozhkov}, {Rozza}, {Sagdeev}, {Sandweiss}, {Saouter},
  {Sbarra}, {Schael}, {Schmidt}, {von Dratzig}, {Schwering}, {Scolieri}, {Seo},
  {Shan}, {Shan}, {Shi}, {Shi}, {Shi}, {Siedenburg}, {Son}, {Spada},
  {Spinella}, {Sun}, {Sun}, {Tacconi}, {Tang}, {Tang}, {Tang}, {Tao},
  {Tescaro}, {Ting}, {Ting}, {Tomassetti}, {Torsti}, {T{\"u}rko{\v{g}}lu},
  {Urban}, {Vagelli}, {Valente}, {Vannini}, {Valtonen}, {Vaurynovich},
  {Vecchi}, {Velasco}, {Vialle}, {Vitale}, {Vitillo}, {Wang}, {Wang}, {Wang},
  {Wang}, {Wang}, {Wang}, {Weng}, {Whitman}, {Wienkenh{\"o}ver}, {Wu}, {Wu},
  {Xia}, {Xie}, {Xie}, {Xiong}, {Xin}, {Xu}, {Xu}, {Yan}, {Yang}, {Yang}, {Ye},
  {Yi}, {Yu}, {Yu}, {Zeissler}, {Zhang}, {Zhang}, {Zhang}, {Zhang}, {Zheng},
  {Zhuang}, {Zhukov}, {Zichichi}, {Zimmermann}, {Zuccon}, {Zurbach}, \& {AMS
  Collaboration}}]{2015PhRvL.114q1103A}
---. 2015{\natexlab{b}}, \prl, 114, 171103

\bibitem[{{Aguilar} {et~al.}(2016){Aguilar}, {Ali Cavasonza}, {Ambrosi},
  {Arruda}, {Attig}, {Aupetit}, {Azzarello}, {Bachlechner}, {Barao}, {Barrau},
  {Barrin}, {Bartoloni}, {Basara}, {Ba{\c{s}}e{\v{g}}mez-du Pree}, {Battarbee},
  {Battiston}, {Becker}, {Behlmann}, {Beischer}, {Berdugo}, {Bertucci},
  {Bindel}, {Bindi}, {Boella}, {de Boer}, {Bollweg}, {Bonnivard}, {Borgia},
  {Boschini}, {Bourquin}, {Bueno}, {Burger}, {Cadoux}, {Cai}, {Capell},
  {Caroff}, {Casaus}, {Castellini}, {Cervelli}, {Chae}, {Chang}, {Chen},
  {Chen}, {Chen}, {Cheng}, {Chou}, {Choumilov}, {Choutko}, {Chung}, {Clark},
  {Clavero}, {Coignet}, {Consolandi}, {Contin}, {Corti}, {Creus},
  {Crispoltoni}, {Cui}, {Dai}, {Delgado}, {Della Torre}, {Demakov},
  {Demirk{\"o}z}, {Derome}, {Di Falco}, {Dimiccoli}, {D{\'\i}az}, {von
  Doetinchem}, {Dong}, {Donnini}, {Duranti}, {D'Urso}, {Egorov}, {Eline},
  {Eronen}, {Feng}, {Fiandrini}, {Finch}, {Fisher}, {Formato}, {Galaktionov},
  {Gallucci}, {Garc{\'\i}a}, {Garc{\'\i}a-L{\'o}pez}, {Gargiulo}, {Gast},
  {Gebauer}, {Gervasi}, {Ghelfi}, {Giovacchini}, {Goglov}, {G{\'o}mez-Coral},
  {Gong}, {Goy}, {Grabski}, {Grandi}, {Graziani}, {Guo}, {Haino}, {Han}, {He},
  {Heil}, {Hoffman}, {Hsieh}, {Huang}, {Huang}, {Huh}, {Incagli}, {Ionica},
  {Jang}, {Jinchi}, {Kang}, {Kanishev}, {Kim}, {Kim}, {Kirn}, {Konak},
  {Kounina}, {Kounine}, {Koutsenko}, {Krafczyk}, {La Vacca}, {Laudi},
  {Laurenti}, {Lazzizzera}, {Lebedev}, {Lee}, {Lee}, {Leluc}, {Li}, {Li}, {Li},
  {Li}, {Li}, {Li}, {Li}, {Li}, {Li}, {Lim}, {Lin}, {Lipari}, {Lippert}, {Liu},
  {Liu}, {Lordello}, {Lu}, {Lu}, {Luebelsmeyer}, {Luo}, {Luo}, {Lv}, {Machate},
  {Majka}, {Ma{\~n}{\'a}}, {Mar{\'\i}n}, {Martin}, {Mart{\'\i}nez}, {Masi},
  {Maurin}, {Menchaca-Rocha}, {Meng}, {Mikuni}, {Mo}, {Morescalchi}, {Mott},
  {Nelson}, {Ni}, {Nikonov}, {Nozzoli}, {Oliva}, {Orcinha}, {Palmonari},
  {Palomares}, {Paniccia}, {Pauluzzi}, {Pensotti}, {Pereira}, {Picot-Clemente},
  {Pilo}, {Pizzolotto}, {Plyaskin}, {Pohl}, {Poireau}, {Putze}, {Quadrani},
  {Qi}, {Qin}, {Qu}, {R{\"a}ih{\"a}}, {Rancoita}, {Rapin}, {Ricol},
  {Rosier-Lees}, {Rozhkov}, {Rozza}, {Sagdeev}, {Sandweiss}, {Saouter},
  {Schael}, {Schmidt}, {Schulz von Dratzig}, {Schwering}, {Seo}, {Shan}, {Shi},
  {Siedenburg}, {Son}, {Song}, {Sun}, {Tacconi}, {Tang}, {Tang}, {Tao},
  {Tescaro}, {Ting}, {Ting}, {Tomassetti}, {Torsti}, {T{\"u}rko{\v{g}}lu},
  {Urban}, {Vagelli}, {Valente}, {Vannini}, {Valtonen}, {V{\'a}zquez Acosta},
  {Vecchi}, {Velasco}, {Vialle}, {Vitale}, {Vitillo}, {Wang}, {Wang}, {Wang},
  {Wang}, {Wang}, {Wang}, {Wei}, {Weng}, {Whitman}, {Wienkenh{\"o}ver}, {Wu},
  {Wu}, {Xia}, {Xiong}, {Xu}, {Yan}, {Yang}, {Yang}, {Yang}, {Yi}, {Yu}, {Yu},
  {Zeissler}, {Zhang}, {Zhang}, {Zhang}, {Zhang}, {Zhang}, {Zhang}, {Zheng},
  {Zhu}, {Zhuang}, {Zhukov}, {Zichichi}, {Zimmermann}, {Zuccon}, \& {AMS
  Collaboration}}]{2016PhRvL.117w1102A}
{Aguilar}, M., {Ali Cavasonza}, L., {Ambrosi}, G., {et~al.} 2016, \prl, 117,
  231102

\bibitem[{{Ahlers}(2016)}]{2016PhRvL.117o1103A}
{Ahlers}, M. 2016, \prl, 117, 151103

\bibitem[{{Ahlers} \& {Mertsch}(2017)}]{2017PrPNP..94..184A}
{Ahlers}, M., \& {Mertsch}, P. 2017, Progress in Particle and Nuclear Physics,
  94, 184

\bibitem[{{Ahn} {et~al.}(2008){Ahn}, {Allison}, {Bagliesi}, {Beatty},
  {Bigongiari}, {Boyle}, {Brandt}, {Childers}, {Conklin}, {Coutu}, {Duvernois},
  {Ganel}, {Han}, {Hyun}, {Jeon}, {Kim}, {Lee}, {Lee}, {Lutz}, {Maestro},
  {Malinin}, {Marrocchesi}, {Minnick}, {Mognet}, {Nam}, {Nutter}, {Park},
  {Park}, {Seo}, {Sina}, {Swordy}, {Wakely}, {Wu}, {Yang}, {Yoon}, {Zei}, \&
  {Zinn}}]{2008APh....30..133A}
{Ahn}, H.~S., {Allison}, P.~S., {Bagliesi}, M.~G., {et~al.} 2008, Astroparticle
  Physics, 30, 133

\bibitem[{{Alemanno} {et~al.}(2021){Alemanno}, {An}, {Azzarello}, {Barbato},
  {Bernardini}, {Bi}, {Cai}, {Catanzani}, {Chang}, {Chen}, {Chen}, {Chen},
  {Cui}, {Cui}, {Cui}, {Dai}, {D'Amone}, {de Benedittis}, {de Mitri}, {de
  Palma}, {Deliyergiyev}, {di Santo}, {Dong}, {Dong}, {Donvito}, {Droz},
  {Duan}, {Duan}, {D'Urso}, {Fan}, {Fan}, {Fang}, {Fang}, {Feng}, {Feng},
  {Fusco}, {Gao}, {Gargano}, {Gong}, {Gong}, {Guo}, {Guo}, {Guo}, {Han}, {Hu},
  {Huang}, {Huang}, {Huang}, {Ionica}, {Jiang}, {Kong}, {Kotenko}, {Kyratzis},
  {Lei}, {Li}, {Li}, {Li}, {Li}, {Liang}, {Liu}, {Liu}, {Liu}, {Liu}, {Liu},
  {Liu}, {Loparco}, {Luo}, {Ma}, {Ma}, {Ma}, {Ma}, {Marsella}, {Mazziotta},
  {Mo}, {Niu}, {Pan}, {Parenti}, {Peng}, {Peng}, {Perrina}, {Qiao}, {Rao},
  {Ruina}, {Salinas}, {Shang}, {Shen}, {Shen}, {Shen}, {Silveri}, {Song},
  {Stolpovskiy}, {Su}, {Su}, {Sun}, {Surdo}, {Teng}, {Tykhonov}, {Wang},
  {Wang}, {Wang}, {Wang}, {Wang}, {Wang}, {Wang}, {Wang}, {Wang}, {Wei}, {Wei},
  {Wei}, {Wen}, {Wu}, {Wu}, {Wu}, {Wu}, {Wu}, {Xia}, {Xu}, {Xu}, {Xu}, {Xu},
  {Xue}, {Yang}, {Yang}, {Yang}, {Yao}, {Yu}, {Yuan}, {Yuan}, {Yue}, {Zang},
  {Zhang}, {Zhang}, {Zhang}, {Zhang}, {Zhang}, {Zhang}, {Zhang}, {Zhang},
  {Zhang}, {Zhang}, {Zhao}, {Zhao}, {Zhao}, {Zhou}, {Zhu}, \& {Dampe
  Collaboration}}]{2021PhRvL.126t1102A}
{Alemanno}, F., {An}, Q., {Azzarello}, P., {et~al.} 2021, \prl, 126, 201102

\bibitem[{{Amenomori} {et~al.}(2005){Amenomori}, {Ayabe}, {Cui},
  {Danzengluobu}, {Ding}, {Ding}, {Feng}, {Feng}, {Gao}, {Geng}, {Guo}, {He},
  {He}, {Hibino}, {Hotta}, {Hu}, {Hu}, {Huang}, {Huang}, {Jia}, {Kajino},
  {Kasahara}, {Katayose}, {Kato}, {Kawata}, {Labaciren}, {Le}, {Li}, {Lu},
  {Lu}, {Meng}, {Mizutani}, {Mu}, {Munakata}, {Nagai}, {Nanjo}, {Nishizawa},
  {Ohnishi}, {Ohta}, {Onuma}, {Ouchi}, {Ozawa}, {Ren}, {Saito}, {Sakata},
  {Sasaki}, {Shibata}, {Shiomi}, {Shirai}, {Sugimoto}, {Takita}, {Tan},
  {Tateyama}, {Torii}, {Tsuchiya}, {Udo}, {Utsugi}, {Wang}, {Wang}, {Wang},
  {Wang}, {Wu}, {Xue}, {Yamamoto}, {Yan}, {Yang}, {Yasue}, {Ye}, {Yu}, {Yuan},
  {Yuda}, {Zhang}, {Zhang}, {Zhang}, {Zhang}, {Zhang}, {Zhang}, {Zhaxisangzhu},
  {Zhou}, \& {Tibet AS{\ensuremath{\gamma}}
  Collaboration}}]{2005ApJ...626L..29A}
{Amenomori}, M., {Ayabe}, S., {Cui}, S.~W., {et~al.} 2005, \apjl, 626, L29

\bibitem[{{Amenomori} {et~al.}(2017){Amenomori}, {Bi}, {Chen}, {Chen}, {Chen},
  {Cui}, {Danzengluobu}, {Ding}, {Feng}, {Feng}, {Feng}, {Gou}, {Guo}, {He},
  {He}, {Hibino}, {Hotta}, {Hu}, {Hu}, {Huang}, {Jia}, {Jiang}, {Kajino},
  {Kasahara}, {Katayose}, {Kato}, {Kawata}, {Kozai}, {Labaciren}, {Le}, {Li},
  {Li}, {Li}, {Liu}, {Liu}, {Liu}, {Lu}, {Meng}, {Miyazaki}, {Mizutani},
  {Munakata}, {Nakajima}, {Nakamura}, {Nanjo}, {Nishizawa}, {Niwa}, {Ohnishi},
  {Ohta}, {Ozawa}, {Qian}, {Qu}, {Saito}, {Saito}, {Sakata}, {Sako}, {Shao},
  {Shibata}, {Shiomi}, {Shirai}, {Sugimoto}, {Takita}, {Tan}, {Tateyama},
  {Torii}, {Tsuchiya}, {Udo}, {Wang}, {Wu}, {Xue}, {Yamamoto}, {Yamauchi},
  {Yang}, {Yuan}, {Yuda}, {Zhai}, {Zhang}, {Zhang}, {Zhang}, {Zhang}, {Zhang},
  {Zhang}, {Zhaxisangzhu}, {Zhou}, \& {Tibet AS{\ensuremath{\gamma}}
  Collaboration}}]{2017ApJ...836..153A}
{Amenomori}, M., {Bi}, X.~J., {Chen}, D., {et~al.} 2017, \apj, 836, 153

\bibitem[{{An} {et~al.}(2019){An}, {Asfandiyarov}, {Azzarello}, {Bernardini},
  {Bi}, {Cai}, {Chang}, {Chen}, {Chen}, {Chen}, {Chen}, {Cui}, {Cui}, {Dai},
  {D'Amone}, {De Benedittis}, {De Mitri}, {Di Santo}, {Ding}, {Dong}, {Dong},
  {Dong}, {Donvito}, {Droz}, {Duan}, {Duan}, {D'Urso}, {Fan}, {Fan}, {Fang},
  {Feng}, {Feng}, {Fusco}, {Gallo}, {Gan}, {Gao}, {Gargano}, {Gong}, {Gong},
  {Guo}, {Guo}, {Guo}, {Han}, {Hu}, {Huang}, {Huang}, {Huang}, {Ionica},
  {Jiang}, {Jin}, {Kong}, {Lei}, {Li}, {Li}, {Li}, {Li}, {Li}, {Liang},
  {Liang}, {Liao}, {Liu}, {Liu}, {Liu}, {Liu}, {Liu}, {Liu}, {Loparco}, {Luo},
  {Ma}, {Ma}, {Ma}, {Ma}, {Ma}, {Marsella}, {Mazziotta}, {Mo}, {Niu}, {Pan},
  {Peng}, {Peng}, {Qiao}, {Rao}, {Salinas}, {Shang}, {Shen}, {Shen}, {Shen},
  {Song}, {Su}, {Su}, {Sun}, {Surdo}, {Teng}, {Tykhonov}, {Vitillo}, {Wang},
  {Wang}, {Wang}, {Wang}, {Wang}, {Wang}, {Wang}, {Wang}, {Wang}, {Wang},
  {Wang}, {Wang}, {Wang}, {Wei}, {Wei}, {Wei}, {Wen}, {Wu}, {Wu}, {Wu}, {Wu},
  {Wu}, {Xi}, {Xia}, {Xu}, {Xu}, {Xu}, {Xu}, {Xue}, {Yang}, {Yang}, {Yang},
  {Yang}, {Yao}, {Yu}, {Yuan}, {Yue}, {Zang}, {Zhang}, {Zhang}, {Zhang},
  {Zhang}, {Zhang}, {Zhang}, {Zhang}, {Zhang}, {Zhang}, {Zhang}, {Zhang},
  {Zhang}, {Zhang}, {Zhao}, {Zhao}, {Zhao}, {Zhou}, {Zhou}, {Zhu}, {Zhu}, \&
  {Zimmer}}]{2019SciA....5.3793A}
{An}, Q., {Asfandiyarov}, R., {Azzarello}, P., {et~al.} 2019, Science Advances,
  5, eaax3793

\bibitem[{{Antoni} {et~al.}(2005){Antoni}, {Apel}, {Badea}, {Bekk}, {Bercuci},
  {Bl{\"u}mer}, {Bozdog}, {Brancus}, {Chilingarian}, {Daumiller}, {Doll},
  {Engel}, {Engler}, {Fe{\ss}ler}, {Gils}, {Glasstetter}, {Haungs}, {Heck},
  {H{\"o}randel}, {Kampert}, {Klages}, {Maier}, {Mathes}, {Mayer}, {Milke},
  {M{\"u}ller}, {Obenland}, {Oehlschl{\"a}ger}, {Ostapchenko}, {Petcu},
  {Rebel}, {Risse}, {Risse}, {Roth}, {Schatz}, {Schieler}, {Scholz}, {Thouw},
  {Ulrich}, {van Buren}, {Vardanyan}, {Weindl}, {Wochele}, \&
  {Zabierowski}}]{2005APh....24....1A}
{Antoni}, T., {Apel}, W.~D., {Badea}, A.~F., {et~al.} 2005, Astroparticle
  Physics, 24, 1

\bibitem[{{Apel} {et~al.}(2019){Apel}, {Arteaga-Vel{\'a}zquez}, {Bekk},
  {Bertaina}, {Bl{\"u}mer}, {Bonino}, {Bozdog}, {Brancus}, {Cantoni},
  {Chiavassa}, {Cossavella}, {Daumiller}, {de Souza}, {Di Pierro}, {Doll},
  {Engel}, {Fuhrmann}, {Gherghel-Lascu}, {Gils}, {Glasstetter}, {Grupen},
  {Haungs}, {Heck}, {H{\"o}randel}, {Huege}, {Kampert}, {Kang}, {Klages},
  {Link}, {{\L}uczak}, {Mathes}, {Mayer}, {Milke}, {Mitrica}, {Morello},
  {Oehlschl{\"a}ger}, {Ostapchenko}, {Pierog}, {Rebel}, {Roth}, {Schieler},
  {Schoo}, {Schr{\"o}der}, {Sima}, {Toma}, {Trinchero}, {Ulrich}, {Weindl},
  {Wochele}, \& {Zabierowski}}]{2019ApJ...870...91A}
{Apel}, W.~D., {Arteaga-Vel{\'a}zquez}, J.~C., {Bekk}, K., {et~al.} 2019, \apj,
  870, 91

\bibitem[{{Atkin} {et~al.}(2018){Atkin}, {Bulatov}, {Dorokhov}, {Gorbunov},
  {Filippov}, {Grebenyuk}, {Karmanov}, {Kovalev}, {Kudryashov}, {Kurganov},
  {Merkin}, {Panov}, {Podorozhny}, {Polkov}, {Porokhovoy}, {Shumikhin},
  {Tkachenko}, {Tkachev}, {Turundaevskiy}, {Vasiliev}, \&
  {Voronin}}]{2018JETPL.108....5A}
{Atkin}, E., {Bulatov}, V., {Dorokhov}, V., {et~al.} 2018, Soviet Journal of
  Experimental and Theoretical Physics Letters, 108, 5

\bibitem[{{Bai} {et~al.}(2019){Bai}, {Bi}, {Bi}, {Cao}, {Chen}, {Chen},
  {Chiavassa}, {Cui}, {Dai}, {della Volpe}, {Di Girolamo}, {Di Sciascio},
  {Fan}, {Giacalone}, {Guo}, {He}, {He}, {Heller}, {Huang}, {Huang}, {Jia},
  {Ksenofontov}, {Leahy}, {Li}, {Li}, {Liang}, {Lipari}, {Liu}, {Liu}, {Liu},
  {Ma}, {Martineau-Huynh}, {Martraire}, {Montaruli}, {Ruffolo}, {Stenkin},
  {Su}, {Tam}, {Tang}, {Tian}, {Vallania}, {Vernetto}, {Vigorito}, {Wang},
  {Wang}, {Wang}, {Wang}, {Wang}, {Wang}, {Wei}, {Wei}, {Wu}, {Wu}, {Wu},
  {Yan}, {Yang}, {Yang}, {Yao}, {Yin}, {Yuan}, {Zhang}, {Zhang}, {Zhang},
  {Zhang}, {Zhang}, {Zhang}, {Zhao}, {Zhou}, {Zhu}, \&
  {Zhu}}]{2019arXiv190502773B}
{Bai}, X., {Bi}, B.~Y., {Bi}, X.~J., {et~al.} 2019, arXiv e-prints,
  arXiv:1905.02773

\bibitem[{{Bartoli} {et~al.}(2015){Bartoli}, {Bernardini}, {Bi}, {Cao},
  {Catalanotti}, {Chen}, {Chen}, {Cui}, {Dai}, {D'Amone}, {Danzengluobu}, {De
  Mitri}, {D'Ettorre Piazzoli}, {Di Girolamo}, {Di Sciascio}, {Feng}, {Feng},
  {Feng}, {Gao}, {Gou}, {Guo}, {He}, {Hu}, {Hu}, {Iacovacci}, {Iuppa}, {Jia},
  {Labaciren}, {Li}, {Liu}, {Liu}, {Liu}, {Lu}, {Ma}, {Ma}, {Mancarella},
  {Mari}, {Marsella}, {Mastroianni}, {Montini}, {Ning}, {Perrone}, {Pistilli},
  {Salvini}, {Santonico}, {Shen}, {Sheng}, {Shi}, {Surdo}, {Tan}, {Vallania},
  {Vernetto}, {Vigorito}, {Wang}, {Wu}, {Wu}, {Xue}, {Yang}, {Yang}, {Yao},
  {Yuan}, {Zha}, {Zhang}, {Zhang}, {Zhang}, {Zhang}, {Zhao}, {Zhaxiciren},
  {Zhaxisangzhu}, {Zhou}, {Zhu}, {Zhu}, \& {ARGO-YBJ
  Collaboration}}]{2015ApJ...809...90B}
{Bartoli}, B., {Bernardini}, P., {Bi}, X.~J., {et~al.} 2015, \apj, 809, 90

\bibitem[{{Bartoli} {et~al.}(2018){Bartoli}, {Bernardini}, {Bi}, {Cao},
  {Catalanotti}, {Chen}, {Chen}, {Cui}, {Dai}, {D'Amone}, {Danzengluobu}, {De
  Mitri}, {D'Ettorre Piazzoli}, {Di Girolamo}, {Di Sciascio}, {Feng}, {Feng},
  {Gao}, {Gou}, {Guo}, {He}, {Hu}, {Hu}, {Iacovacci}, {Iuppa}, {Jia},
  {Labaciren}, {Li}, {Liu}, {Liu}, {Liu}, {Lu}, {Ma}, {Ma}, {Mancarella},
  {Mari}, {Marsella}, {Mastroianni}, {Montini}, {Ning}, {Perrone}, {Pistilli},
  {Ruffolo}, {Salvini}, {Santonico}, {Shen}, {Sheng}, {Shi}, {Surdo}, {Tan},
  {Vallania}, {Vernetto}, {Vigorito}, {Wang}, {Wu}, {Wu}, {Xue}, {Yang},
  {Yang}, {Yao}, {Yuan}, {Zha}, {Zhang}, {Zhang}, {Zhang}, {Zhang}, {Zhao},
  {Zhaxiciren}, {Zhaxisangzhu}, {Zhou}, {Zhu}, {Zhu}, \& {ARGO-YBJ
  Collaboration}}]{2018ApJ...861...93B}
---. 2018, \apj, 861, 93

\bibitem[{{Blasi} {et~al.}(2012){Blasi}, {Amato}, \&
  {Serpico}}]{2012PhRvL.109f1101B}
{Blasi}, P., {Amato}, E., \& {Serpico}, P.~D. 2012, \prl, 109, 061101

\bibitem[{{Engelbrecht} {et~al.}(2017){Engelbrecht}, {Strauss}, {le Roux}, \&
  {Burger}}]{2017ApJ...841..107E}
{Engelbrecht}, N.~E., {Strauss}, R.~D., {le Roux}, J.~A., \& {Burger}, R.~A.
  2017, \apj, 841, 107

\bibitem[{{Faherty} {et~al.}(2007){Faherty}, {Walter}, \&
  {Anderson}}]{2007Ap&SS.308..225F}
{Faherty}, J., {Walter}, F.~M., \& {Anderson}, J. 2007, \apss, 308, 225

\bibitem[{{Forman} \& {Gleeson}(1975)}]{1975Ap&SS..32...77F}
{Forman}, M.~A., \& {Gleeson}, L.~J. 1975, \apss, 32, 77

\bibitem[{{Fornieri} {et~al.}(2020){Fornieri}, {Gaggero}, {Guberman},
  {Brahimi}, \& {Marcowith}}]{2020arXiv200715321F}
{Fornieri}, O., {Gaggero}, D., {Guberman}, D., {Brahimi}, L., \& {Marcowith},
  A. 2020, arXiv e-prints, arXiv:2007.15321

\bibitem[{{Frisch} {et~al.}(2015){Frisch}, {Berdyugin}, {Piirola}, {Magalhaes},
  {Seriacopi}, {Wiktorowicz}, {Andersson}, {Funsten}, {McComas}, {Schwadron},
  {Slavin}, {Hanson}, \& {Fu}}]{2015ApJ...814..112F}
{Frisch}, P.~C., {Berdyugin}, A., {Piirola}, V., {et~al.} 2015, \apj, 814, 112

\bibitem[{{Funsten} {et~al.}(2013){Funsten}, {DeMajistre}, {Frisch},
  {Heerikhuisen}, {Higdon}, {Janzen}, {Larsen}, {Livadiotis}, {McComas},
  {M{\"o}bius}, {Reese}, {Reisenfeld}, {Schwadron}, \&
  {Zirnstein}}]{2013ApJ...776...30F}
{Funsten}, H.~O., {DeMajistre}, R., {Frisch}, P.~C., {et~al.} 2013, \apj, 776,
  30

\bibitem[{{Gleeson} \& {Axford}(1967)}]{1967ApJ...149L.115G}
{Gleeson}, L.~J., \& {Axford}, W.~I. 1967, \apjl, 149, L115

\bibitem[{{Gleeson} \& {Axford}(1968)}]{1968ApJ...154.1011G}
---. 1968, \apj, 154, 1011

\bibitem[{{Liu} {et~al.}(2019){Liu}, {Yan}, \& {Zhang}}]{2019PhRvL.123v1103L}
{Liu}, R.-Y., {Yan}, H., \& {Zhang}, H. 2019, \prl, 123, 221103

\bibitem[{{Liu} {et~al.}(2017){Liu}, {Bi}, {Lin}, {Wang}, \&
  {Yin}}]{2017PhRvD..96b3006L}
{Liu}, W., {Bi}, X.-J., {Lin}, S.-J., {Wang}, B.-B., \& {Yin}, P.-F. 2017,
  \prd, 96, 023006

\bibitem[{{Liu} {et~al.}(2018){Liu}, {Yao}, \& {Guo}}]{2018ApJ...869..176L}
{Liu}, W., {Yao}, Y.-h., \& {Guo}, Y.-Q. 2018, \apj, 869, 176

\bibitem[{{Malkov} {et~al.}(2012){Malkov}, {Diamond}, \&
  {Sagdeev}}]{2012PhRvL.108h1104M}
{Malkov}, M.~A., {Diamond}, P.~H., \& {Sagdeev}, R.~Z. 2012, \prl, 108, 081104

\bibitem[{{Mou} {et~al.}(2018){Mou}, {Sun}, \& {Xie}}]{2018ApJ...869L..20M}
{Mou}, G., {Sun}, D., \& {Xie}, F. 2018, \apjl, 869, L20

\bibitem[{{Ohira} \& {Ioka}(2011)}]{2011ApJ...729L..13O}
{Ohira}, Y., \& {Ioka}, K. 2011, \apjl, 729, L13

\bibitem[{{Ptuskin} {et~al.}(2013){Ptuskin}, {Zirakashvili}, \&
  {Seo}}]{2013ApJ...763...47P}
{Ptuskin}, V., {Zirakashvili}, V., \& {Seo}, E.-S. 2013, \apj, 763, 47

\bibitem[{{Qiao} {et~al.}(2019){Qiao}, {Liu}, {Guo}, \&
  {Yuan}}]{2019JCAP...12..007Q}
{Qiao}, B.-Q., {Liu}, W., {Guo}, Y.-Q., \& {Yuan}, Q. 2019, \jcap, 2019, 007

\bibitem[{{Salvati} \& {Sacco}(2008)}]{2008A&A...485..527S}
{Salvati}, M., \& {Sacco}, B. 2008, \aap, 485, 527

\bibitem[{{Taylor} \& {Giacinti}(2017)}]{2017PhRvD..95b3001T}
{Taylor}, A.~M., \& {Giacinti}, G. 2017, \prd, 95, 023001

\bibitem[{{Thoudam} \& {H{\"o}randel}(2014)}]{2014A&A...567A..33T}
{Thoudam}, S., \& {H{\"o}randel}, J.~R. 2014, \aap, 567, A33

\bibitem[{{Tibaldo} {et~al.}(2015){Tibaldo}, {Digel}, {Casandjian},
  {Franckowiak}, {Grenier}, {J{\'o}hannesson}, {Marshall}, {Moskalenko},
  {Negro}, {Orlando}, {Porter}, {Reimer}, \& {Strong}}]{2015ApJ...807..161T}
{Tibaldo}, L., {Digel}, S.~W., {Casandjian}, J.~M., {et~al.} 2015, \apj, 807,
  161

\bibitem[{{Tibet AS{\ensuremath{\gamma}} Collaboration} {et~al.}(2006){Tibet
  AS{\ensuremath{\gamma}} Collaboration}, {Amenomori}, {Ayabe}, {Chen}, {Cui},
  {Danzengluobu}, {Ding}, {Ding}, {Feng}, {Feng}, {Gao}, {Geng}, {Guo}, {He},
  {He}, {Hibino}, {Hotta}, {Hu}, {Hu}, {Huang}, {Huang}, {Jia}, {Kajino},
  {Kasahara}, {Katayose}, {Kato}, {Kawata}, {Labaciren}, {Le}, {Li}, {Lu},
  {Lu}, {Meng}, {Mizutani}, {Mu}, {Munakata}, {Nagai}, {Nanjo}, {Nishizawa},
  {Ohnishi}, {Ohta}, {Onuma}, {Ouchi}, {Ozawa}, {Ren}, {Saito}, {Sakata},
  {Sasaki}, {Shibata}, {Shiomi}, {Shirai}, {Sugimoto}, {Takita}, {Tan},
  {Tateyama}, {Torii}, {Tsuchiya}, {Udo}, {Wang}, {Wang}, {Wang}, {Wu}, {Xue},
  {Yamamoto}, {Yan}, {Yang}, {Yasue}, {Ye}, {Yu}, {Yuan}, {Yuda}, {Zhang},
  {Zhang}, {Zhang}, {Zhang}, {Zhang}, {Zhang}, {Zhaxisangzhu}, \&
  {Zhou}}]{2006PhLB..632...58T}
{Tibet AS{\ensuremath{\gamma}} Collaboration}, {Amenomori}, M., {Ayabe}, S.,
  {et~al.} 2006, Physics Letters B, 632, 58

\bibitem[{{Tomassetti}(2012)}]{2012ApJ...752L..13T}
{Tomassetti}, N. 2012, \apjl, 752, L13

\bibitem[{{Tomassetti}(2015)}]{2015ApJ...815L...1T}
---. 2015, \apjl, 815, L1

\bibitem[{{Yoon} {et~al.}(2017){Yoon}, {Anderson}, {Barrau}, {Conklin},
  {Coutu}, {Derome}, {Han}, {Jeon}, {Kim}, {Kim}, {Lee}, {Lee}, {Lee}, {Lee},
  {Link}, {Menchaca-Rocha}, {Mitchell}, {Mognet}, {Nutter}, {Park},
  {Picot-Clemente}, {Putze}, {Seo}, {Smith}, \& {Wu}}]{2017ApJ...839....5Y}
{Yoon}, Y.~S., {Anderson}, T., {Barrau}, A., {et~al.} 2017, \apj, 839, 5

\bibitem[{{Yuan} {et~al.}(2011){Yuan}, {Zhang}, \& {Bi}}]{2011PhRvD..84d3002Y}
{Yuan}, Q., {Zhang}, B., \& {Bi}, X.-J. 2011, \prd, 84, 043002

\bibitem[{{Yue} {et~al.}(2019){Yue}, {Ma}, {Yuan}, {Fan}, {Chen}, {Cui}, {Dai},
  {Dong}, {Huang}, {Jiang}, {Lei}, {Li}, {Liu}, {Liu}, {Liu}, {Luo}, {Pan},
  {Peng}, {Qiao}, {Wei}, {Wu}, {Xu}, {Xu}, {Yuan}, {Zang}, {Zhang}, {Zhang}, \&
  {Zhang}}]{2019FrPhy..1524601Y}
{Yue}, C., {Ma}, P.-X., {Yuan}, Q., {et~al.} 2019, Frontiers of Physics, 15,
  24601

\bibitem[{{Zhang} {et~al.}(2017){Zhang}, {Liu}, \&
  {Yuan}}]{2017ApJ...844L...3Z}
{Zhang}, Y., {Liu}, S., \& {Yuan}, Q. 2017, \apjl, 844, L3

\bibitem[{{Zirnstein} {et~al.}(2016){Zirnstein}, {Heerikhuisen}, {Funsten},
  {Livadiotis}, {McComas}, \& {Pogorelov}}]{2016ApJ...818L..18Z}
{Zirnstein}, E.~J., {Heerikhuisen}, J., {Funsten}, H.~O., {et~al.} 2016, \apjl,
  818, L18

\end{thebibliography}
\begin{table}[b]
	\centering
	\begin{tabular}{c|ccc}
		 & Fig.~\ref{f1} & Fig.~\ref{f2} & Unit\\
		\hline
		$ \chi \left( \text{GV}\right) $ & \multicolumn{2}{c}{0.15} & $ 1 $\\
		$ \alpha $ & \multicolumn{2}{c}{2.6} & 1\\
		$ R_\text{m} $ & \multicolumn{2}{c}{19} & PV\\
		$ \phi $ & \multicolumn{2}{c}{970} & MV\\
		$ \theta $ & $ 5.5 $ & $ 4.7 $ & deg\\
		$ \varphi $ & NA & 90 & deg\\
		$ M_\text{A} $ & 0.11 & 0.1 & 1\\
		$ \left| z/h \right| $ & \multicolumn{2}{c}{0.3} & $ 1 $\\
		$ \mu Q_{\text{d}}\left( \text{GV}\right) $ & \multicolumn{2}{c}{$ 7\times 10^{53} $} & $ \text{kpc}^{-2}\text{Myr}^{-1}\text{GV}^{-1} $\\
		$ Q_{\text{c}}\left( \text{GV}\right) $ & $ 2.1\times 10^{57} $ & $ 8.8\times 10^{56} $ & $ \text{Myr}^{-1}\text{GV}^{-1} $\\
		$ t_{\text{c}} $ & \multicolumn{2}{c}{18} & Myr
	\end{tabular}
	\caption{Fitting parameters for models of Figs.~\ref{f1} and \ref{f2}.}\label{t1}
\end{table}
\end{document}